\RequirePackage{fix-cm}
\documentclass[natbib]{svjour3}                     
\smartqed  

\usepackage{graphicx}
\usepackage{color}
\usepackage{graphics}
\usepackage{multirow}
\usepackage{array,ragged2e}
\usepackage{url}
\usepackage{subfigure}
\usepackage{marvosym}
\usepackage{longtable}
\usepackage{lscape}
\usepackage[counterclockwise]{rotating} 

\newcounter{findings}

\newcommand{\newfinding}[4]{\refstepcounter{findings}\textbf{F\arabic{findings})} #4 \label{find:#1#2#3}\label{find:\arabic{findings}}}

\newcolumntype{C}[1]{>{\Centering\arraybackslash\hspace{0pt}}m{#1}}
\newcolumntype{R}[1]{>{\RaggedLeft\arraybackslash\hspace{0pt}}m{#1}}
\newcolumntype{L}[1]{>{\RaggedRight\arraybackslash\hspace{0pt}}m{#1}}

\newcommand{\rb}[1]{\rotatebox{-90}{\textit{#1}}}

\newcommand{\findref}[3]{F\ref{find:#1#2#3}}

\newenvironment{procedureList}
{\renewcommand{\labelenumi}{\Roman{enumi}.}
}
{\renewcommand{\labelenumi}{\arabic{enumi})}
}

\spnewtheorem{rquestion}{RQ}{\bf}{\it}

\journalname{Software Quality Journal}

\begin{document}

\title{
Introduction of Static Quality Analysis
in Small and Medium-Sized Software Enterprises:
Experiences from Technology Transfer
}
\titlerunning{Static Quality Analysis in SMEs}        

\author{
Mario Gleirscher \and 
Dmitriy Golubitskiy \and 
Maximilian Irlbeck \and
Stefan Wagner}

\authorrunning{M.~Gleirscher, D.~Golubitskiy, M.~Irlbeck, S.~Wagner}

\institute{M.~Gleirscher, M.~Irlbeck \at 
Institut f\"ur Informatik, Technische Universit\"at M\"unchen, 
Germany\\
\email{{gleirsch,irlbeck}@in.tum.de}
\and
S.~Wagner \at
Institute of Software Technology, University of Stuttgart, Germany\\
\email{Stefan.Wagner@informatik.uni-stuttgart.de}
\and
D.~Golubitskiy \at
Roland Berger Strategy Consultants GmbH, M\"unchen, Germany\\
\email{Dmitriy.Golubitskiy@rolandberger.com}
}

\index{Gleirscher, Mario} 
\index{Golubitskiy, Dmitriy} 
\index{Irlbeck, Maximilian}
\index{Wagner, Stefan}

\date{Received: ? / Accepted: ?}

\maketitle

\begin{abstract}
Today, small and medium-sized enterprises (SMEs) in the software industry face major challenges. Their resource constraints require high efficiency in development. Furthermore, quality assurance (QA) measures need to be taken to mitigate 
the risk of additional, expensive effort for bug fixes or compensations. Automated static analysis (ASA) can reduce this risk because it promises low application effort. 
SMEs seem to take little advantage of this opportunity. Instead, they still mainly rely on the dynamic analysis approach of \textit{software testing}.

In this article, we report on our experiences from a technology transfer project. Our aim was to evaluate the results
static analysis can provide for SMEs as well as the problems that occur when introducing and using static analysis in SMEs.
We analysed five software projects from five collaborating SMEs using three different ASA techniques: code 
clone detection, bug pattern detection and architecture conformance analysis. Following the analysis, we applied a quality
model to aggregate and evaluate the results.

Our study shows that the effort required to introduce ASA techniques in SMEs is small (mostly below one person-hour each).
Furthermore, we encountered only few technical problems. By means of the analyses, we could detect multiple defects in production code.  
The participating companies perceived the analysis results to be a helpful addition to their current QA and will include the analyses in their 
QA process. With the help of the Quamoco quality 
model, we could efficiently aggregate and rate static analysis results. However, we also encountered a partial mismatch with the
opinions of the SMEs.
We conclude, that ASA and quality models can be a valuable and affordable addition to the QA process of SMEs.
\end{abstract}
\vspace{.4cm}
\noindent
\textbf{The final publication is available at \url{http://link.springer.com}.}

\section{Introduction}
\label{sec:intro}

Small and medium-sized enterprises (SMEs) play an important role in the global software industry. In 
many countries, such as the US, Brazil or China, these companies represent up to 85\% of all software 
organisations \citep{Richardson2007} and carry out the majority of software development 
\citep{MISHRA}. Compared to large corporations, SMEs are faced with special conditions such 
as limited resources, lack of expertise or financial insecurity. 

\paragraph{Problem statement.}
While there are many articles focusing on process improvement in SMEs \citep{Kautz1999,MISHRA,Wangenheim2006}, we found no study that considers specific quality assurance (QA) techniques or quality models and their application in this context. However, QA constitutes an important and resource intensive activity. Automated static analysis (ASA) techniques and associated quality models seem to be suitable for SMEs. These companies usually do not have dedicated quality assurance departments and, therefore, could benefit from highly automated, pre-packaged techniques. The benefits of such techniques lie in their low-cost application \citep{Baca2008} and their potential to detect critical quality defects \citep{Zheng06,ayewah07}. Such defects are risky for further development and increase costs as they might for example entail effort for bug fixes. Detecting these defects at a low cost approach is a promising way for small software enterprises to implement efficient quality assurance.

\paragraph{Research Objective.} Our overall objective is to improve the quality assurance processes at SMEs in a way that suits their specific context. In this article, we focus on the question whether SMEs can benefit from the quality assurance paradigm \emph{static analysis} and associated quality models. Is it possible to introduce ASA techniques into their existing projects with low effort? What kind of defects can be found using these techniques? Is the perceived usefulness for the enterprises strong enough to justify the needed effort?  Finally, can we employ quality models to help in the interpretation of the analyses for an explicit understanding of quality? Answering these questions will support decision makers in SMEs in planning quality assurance improvements in the future.

\paragraph{Contribution.}
We report on our experience in analysing five projects of five SMEs using three different ASA techniques: \emph{code clone detection}, \emph{bug pattern detection} and \emph{architecture conformance analysis}. We evaluate the effort that is needed to introduce these techniques, the pitfalls we came across and how the participating enterprises evaluated the presented techniques as well as the defects we discovered in their software projects. Furthermore, we illustrate our experience in enhancing the static analyses by applying the Quamoco quality model, explain where it confirmed our results and compare its results to the opinions of our study participants.

\paragraph{Results.}
Our study reveals that the effort needed to introduce and apply static analysis techniques is low and affordable for SMEs. Using these techniques, we found critical defects even in production code. After the presentation of the results, all participating SMEs plan to apply the presented techniques in the future. Our study showed that a quality model provides help to interpret the vast amount of findings gained from static analyses. Nevertheless we suggest to pay attention to extreme measures apart from the quality model results. Our study shows that there is a mismatch between  the ratings provided by the quality model and the opinion of the study participants concerning many quality characteristics. 

\paragraph{Outline.}
In Section~\ref{sec:approach}, we describe the research context of our work, 
present our guiding research questions, give a short overview of the applied techniques and models
and explain the procedure we chose. Section~\ref{sec:results} displays the results of our work which we 
discuss in detail in Section~\ref{sec:discussion}. The possible threats to validity of our work are illustrated in Section~\ref{sec:threats}.
Relations of the presented work to other approaches are presented in Section~\ref{sec:relatedwork}.
We conclude our report in Section~\ref{sec:conclusions} and share our perspective on future research.

\section{Approach}
\label{sec:approach}

We describe our experiences of transferring ASA technology to SMEs and its use 
in quality models. This section illustrates the
research context, i.e.~the participating enterprises, our guiding research questions, the 
examined techniques and models, the study objects we employed to gather 
our experiences and, finally, the procedure we used to answer 
our research questions.

\subsection{Research Context}
\label{sec:context}

The basis of our research was the collaboration with five SMEs, 
all resident in the Munich area and selected through personal contacts and a 
series of information events and workshops. Details regarding the selection 
process can be found in Section~\ref{sec:proc}.
Following the definition of the \cite{EuComm2003}, one 
of the participating enterprises is micro-, two are small- and 
two are medium-sized considering their number of staff and annual turnover.
The presented research is based on the experience with these enterprises 
gathered in a project from March 2010 to April 2011.

\subsection{Study Subjects and Objects}
\label{sec:partners}

\paragraph{Study Subjects (SS).}
For our investigation, we collaborate with five SMEs (``study subjects''). These companies 
cover various business and technology domains, e.g.~corporate and local government controlling, form letter processing as 
well as diagnosis and maintenance of embedded systems. Four of them are involved in commercial software 
development, one in software quality assurance and consulting. The latter could not provide a software 
system of their own development.

For each company, we involve one \emph{study participant (SP)} who assumes three roles: (1) our SME contact 
and representative, (2) a stakeholder, executive manager, project leader, supervisor or developer responsible 
for or deeply interested in a study object and, finally, (3) an industrial partner supportively participating in our 
research project. Throughout our study, the SPs are assumed to be partially accompanied by other assistants or SME staff.

\paragraph{Study Objects (SO).}
Our study objects consist of five software systems briefly described in 
Table~\ref{fig:studyobjects}. 
These systems contain between 100 and 600 kLoC. The SOs~1, 2, 4 and 5 have been developed in software projects by the corresponding SPs. 
One study participant, however, did not offer a software system of his own. Following his suggestion, we 
chose the humanitarian open-source application OpenMRS\footnote{\url{http://www.openmrs.org}} to be SO 3. 
This system is a development of the multi-institution, non-profit collaborative OpenMRS. For this SP, we 
additionally consulted the OpenMRS core developers for technical questions. 

The development of all systems started seven 
years  at the earliest before we conducted our study in 2010. The SO project teams comprise less than ten 
persons. Except for OpenMRS, the teams are located in the Munich area. The development of SO~1 and 2 had 
already been finished the start of our study.

\begin{table}[t]
\begin{center}
\renewcommand{\arraystretch}{1.2}	
  \begin{tabular*}{\textwidth}{c|c|c|c|@{\extracolsep{\fill}}p{4.2cm}}
    \textbf{SO} & \textbf{Platform} & \textbf{Sources} & \textbf{Size [kLoC]} & \textbf{Business Domain} \\\hline\hline
    1 & C\#.NET & closed, commercial 	& $\approx 100$ & Corporate controlling \\  

    2 & C\#.NET & closed, commercial 	& $\approx 200$ & Embedded device maintenance \\ 

    3 & Java 	& open, non-profit 		& $\approx 200$ & Health information management\\ 

    4 & Java 	& closed, commercial 	& $\approx 100$ & Local government controlling\\ 

    5 & Java 	& closed, commercial 	& $\approx 560$ & Document processing\\\hline
  \end{tabular*} 
  \caption{Overview of study objects (SO). \textbf{Legend:} kLoC \dots thousand lines of code \label{fig:studyobjects}}
\end{center}
\vspace{-10pt}
\end{table}

\subsection{Research Questions}
\label{sec:questions}

Our overall research objective is to evaluate the transfer of innovative quality assurance measures to SMEs. We structure this objective into three major research questions. The first two analyse the benefit of static analysis techniques, the third
one explores the use of quality models:

\begin{rquestion}
Which problems occur while introducing and applying static analysis techniques at SMEs?
\end{rquestion}

SMEs show special characteristics, such as 
generalist employees instead of specialists for quality assurance. Hence,
it needs to be simple and straightforward to introduce and apply static analysis to be useful
for SMEs. We further break this down into
two sub-questions and indicate our expectations:

\noindent
\textbf{RQ 1.1} \emph{Which technical problems occur?}

Static analysis is tightly coupled to tools that perform and report the
analysis. Hence, the ease of introducing and applying static analysis also
depends on how many and which technical problems the software engineers
need to solve.

\noindent
\textbf{RQ 1.2} \emph{How much effort is necessary?}

If the effort to set up the analyses is too large, 
SME will rule out their application as they cannot afford 
to allocate additional capacities on QA. Therefore, we analyse the effort spent 
on the introduction and application.

\textbf{Expectations for RQ~1:} From our experience, we expect several technical problems
with the configuration of the tools as well as the preparation of the code bases
of the SMEs as the analyses require source code as well as executables. The introduction effort
should not take more than a few person-days per analysed project.

\begin{rquestion}
How useful are automated static analysis techniques for SMEs?
\end{rquestion}

Beyond the question of how easy or problematic it is to introduce and apply static
analyses in SMEs, we are interested whether the techniques can
produce useful results for the developers and quality engineers. Even a small effort should not be
spent if there is no return on investment. Again, we break this
question down into two sub-questions and define our expectations:

\noindent
\textbf{RQ 2.1} \emph{Which defects can be found?}

We investigate usefulness by analysing
the types and numbers of defects found by using the static analysis
tools at the SMEs. If critical defects can be found, the application 
of the techniques is considered useful. We neither focus on specification defects and whether they can be found at all, nor do
we perform cause-and-effects analyses for defects except for some criticality assessments.

\noindent
\textbf{RQ 2.2} \emph{How do the companies perceive the usefulness?}

Motivated by metrics discussed in \cite{Davis1989}, 
we add the subjective perception of our study participants.
How do they interpret the results of the static analysis tools? Do they believe they can work
with those tools and are they going to apply them in their future projects? This way, we augment the information 
we gained from defect analysis.

\textbf{Expectations for RQ~2:} We expect automated static analysis to be useful for SMEs
because of the low effort required for set-up and execution. Most problems found will probably
be related to maintainability, but we also expect to uncover some critical defects.
Overall, we presume that the companies will have a positive impression of the techniques.

\begin{rquestion}
To what degree do the results of ASA, study participants and quality models match?
\end{rquestion}

Findings generated by ASA techniques have to be interpreted in the context 
of overall software quality. This perspective shows to SMEs whether their software
products fulfil overall quality requirements and were these products can be improved. 
It is unclear whether quality models help SMEs to estimate their projects' quality, reveal 
general deficiencies apart from singular findings and 
if their application is worth the effort. We will describe two sub-questions together with our expectations:

\noindent
\textbf{RQ 3.1} \emph{Are individual ASA results well reflected in the quality model results?}

Static analyses can be used to estimate the quality of a certain software product. 
High clone rates or a considerable number of bug pattern findings indicate 
poor software quality. Nevertheless, too many findings can be confusing. We want to know whether the outcomes of the applied
ASA techniques are similar to the results of the Quamoco quality model and if the quality model provides further insights at less complexity.

\noindent
\textbf{RQ 3.2} \emph{Do the results of the quality model match the opinions of the study participants?}

Quality models allow quality engineers to transform findings into ratings of different quality characteristics. Nevertheless, each company develops a
self-perception of their product concerning these characteristics,
which is normally based on an intuitive understanding of these attributes. We want to find out whether
a quality model confirms the self-estimation of SMEs and whether it offers different or additional insights. 

\textbf{Expectations for RQ~3:} We suppose that the quality model is calibrated well enough for the individual results
of ASA to be visible in the assessment results of the model. As suggested by previous studies~\citep{wagner:tse-quamoco},
we assume that there will be a good match between the study participants' opinions and the assessment results.

\renewcommand\theenumi{\arabic{enumi}}
\renewcommand\labelenumi{\theenumi)}

\subsection{Static Analysis Techniques}
\label{sec:methods}

\emph{Static analysis} refers to the analysis of computer programs without their execution. It includes manual techniques, such as
reviews and inspections, as well as automated techniques.
We use the term \emph{static quality analysis} to emphasise the understanding of static analysis results from the perspective 
of software quality attributes.
Manual analyses are time-consuming and prone to missing problems in the
huge amount of code to be analysed. Automation has high potential
to detect simple and reoccurring problems in source code. A detection of the correct usage of ``==''
instead of ``equals'' to compare strings in Java, should not be the task
of human reviewers. They should concentrate on the more subtle and domain-related problems. 

From the interviews with our study participants
and the experiences at our research groups, we chose
three important types of techniques and specialised tools which we introduce in detail
below. Technically, we employ the open-source tool ConQAT\footnote{\url{http://www.conqat.org}} for code clone 
detection and architecture conformance analysis as well as for results processing of bug pattern detection.

\paragraph{Code Clone Detection.}
\label{sec:methods_clones}

Modern programming languages, particularly object-orien\-ted ones, offer various abstraction mechanisms to facilitate reuse of code fragments but copy-paste is still a widely employed reuse strategy. This often leads to numerous duplicated code fragments -- so-called \emph{clones} -- in software systems. As stated in the surveys of \cite{Koschke2007} and \cite{Roy2007}, cloning is problematic for software quality for several reasons:

\begin{itemize}
\item Cloning unnecessarily increases program size and thus efforts for size-related activities like inspections and testing.
\item Changes, including bug fixes, to one clone instance often need to be made to the other instances as well, again increasing efforts.
\item Changes performed inconsistently to duplicated source code fragments can introduce bugs.
\end{itemize}

Code clone detection is an automated static analysis technique that detects duplicated code fragments. 
One of the most important metrics offered by this technique is \textit{unit coverage} which is the probability that 
an arbitrarily chosen \emph{unit} of the source code is part of a clone. 
A unit represents an uncommented and normalised source code statement which originally may have spanned several text 
lines.
Another metric called \textit{blow-up} denotes the ratio of the unit count of the current software by the unit 
count of a hypothetical software without clones \citep{juergens2010achieving}. Moreover, two terms are important for 
clone detection: A \textit{clone class} defines a set of similar code fragments and a \textit{clone instance} is a 
representative of a clone class \citep{Juergens2009}.

We differentiate between \textit{conventional clone detection} and \textit{gapped clone detection}. During conventional clone 
detection, clones are considered to be syntactically similar copies; only variable, type or function identifiers can 
be changed \citep{Koschke2007}. In contrast, gapped clone detection reveals clones with further modifications; 
statements can be changed, added or removed \citep{Koschke2007}. While clones are an indicator of bad design, the 
difference between the two approaches is that only the results of gapped clone detection can reveal defects that 
lead to failures, which arise by unconscious, inconsistent changes in instances of a clone class. We use both 
approaches in our study to investigate both aspects of clone detection.

Clone detection is supported by a number of freely available and commercial tools. The most popular of them are 
CCFinder\footnote{\url{http://www.ccfinder.net}}, 
ConQAT, 
CloneDR\footnote{\url{http://www.semanticdesigns.com/Products/Clone}},
and Axivion Bauhaus Suite\footnote{\url{http://www.axivion.com}}. The former two are free, while the latter two are 
commercial.
We employ ConQAT in our investigation because our research group has experience with its
usage.

\paragraph{Bug Pattern Detection.}
\label{sec:methods_patterns}

By this term we refer to a technique for the automated detection of a variety of defects. 
Bug patterns have been thoroughly investigated, e.g.~by \citet{Zheng06}, and compared with other frequently used software quality assurance techniques such as code reviews or testing \citep{wagner:testcom05}. 
Bug patterns represent a scalable approach to efficiently reveal defects or possible causes thereof. According to \citet{DBLP:conf/icst/WagnerDAWS08}, this technique can already be cost-efficient by detecting three defects that would not have appeared otherwise until system use. 
Their \textit{rules} aim at structural patterns recognisable from source code, executables and meta-data such as source code comments and debug symbols to gain as much knowledge as possible from a static perspective. This knowledge encompasses obvious bugs, rather complex heuristics for latent defects, e.g.~code clones, and less critical issues of coding style.
For example, \emph{uncallable method defined in anonymous class} or \emph{never called} are among the frequently activated rules for Java. These are triggered whenever the applicable code fulfils these conditions.

Because of the large variety of defects, as classified by \citet{Beizer1990:SoftwareTestingTechniques}, there is no complete classification schema for bug patterns yet. A reason for this might be that generally applicable defect classifications 
are rare, vague or difficult to use in practice \citep{DBLP:conf/issta/Wagner08}.
The tools we apply classify their rules according to the consequences of findings such as security vulnerability, performance loss or functional incorrectness. By the term \textit{finding} we denote that a rule was triggered at a specific location. Findings themselves are often categorised by their severity and their confidence levels.

Many of the rules are realised by means of individual lexers and parsers, by using compiler infrastructures or by more reusable means such as pattern or rule languages and machine learning. Rules for latent defects and coding style often stem from abstract source code metrics as discussed, e.g., by \citet{DBLP:conf/iwsm/FerzundAW08}. Among the wide variety of tools%
\footnote{\url{http://en.wikipedia.org/wiki/List_of_tools_for_static_code_analysis}} available for bug pattern detection, are
splint\footnote{\url{http://splint.org}} for C, 
cppcheck\footnote{\url{http://cppcheck.sourceforge.net}} for C++, 
FindBugs\footnote{\url{http://findbugs.sourceforge.net}} \citep{10.1109/MS.2008.130} for Java,
FxCop\footnote{\url{http://msdn.microsoft.com/en-us/library/bb429476\%28v=vs.80\%29.aspx}} for C\# or Coverity Static Analysis for all of these languages \citep{Bessey2010}.
In our study, we use the free tools FxCop and Gendarme for C\# and PMD and FindBugs for Java.

\paragraph{Architecture Conformance.}
\label{sec:methods_archconf}

Architectural erosion denotes the problem that structural knowledge of a system often is lost over its lifetime~\citep{Feilkas2009loss,Fiutem1998identifying,Rosik2008industrial}. 
The documented and the
implemented architectures are drifting
apart. This effect leads to a  decrease in system maintainability. To counteract this, 
different approaches are in use to compare 
the conformance of a system's implementation with its intended architecture.

\citet{passos2010} identify three static concepts existing for architecture conformance analysis:
 \textit{Reflexion Models (RM)}, \textit{Source Code Query Languages (SCQL)} and \textit{Dependency Structure 
Matrices (DSM)}. 

\textit{Reflexion Models}  \citep{Koschke2003} compare two models of a system and assess their conformance. The first 
model usually represents the intended architecture, the second one the implementation of the 
system~\citep{Knodel2007comparison}. The intended architecture consists of components and allowed relationships 
between components, expressed as rules. Each component itself can contain sub-components. The system's source code is 
mapped to these components and then analysed for conformance to the given rules. This technique is used by the 
commercial tools SonarJ\footnote{\url{http://www.hello2morrow.com/products/sonarj}} and 
Structure101\footnote{\url{http://www.headwaysoftware.com}} as well as the open-source tools ConQAT and   
dependometer\footnote{\url{http://source.valtech.com/display/dpm/Dependometer}}.

There are tools using \textit{SCQL} like Semmle.QL~\citep{DeMoor2007QL} or \textit{DSM} like 
Lattix~\citep{Sangal2005Lattix} not further explained here. Both of these concepts rely strongly on the realisation of the 
system and cannot provide 
an architecture specification that is independent of the system's implementation~\citep{Deissenboeck2010}.

Apart from this technical perspective, the quality of an architecture specification is a crucial 
factor for the success of architecture conformance. 
Imprecise, inconsistent, invalid, outdated, coarse-granular or incomplete specifications also decrease the quality 
of findings produced by architecture conformance analysis and hinder the effective creation of 
reflexion models.
Moreover, the strong connection of architecture and system rationales and quality makes it risky to perform architecture conformance analysis on the basis of a defective architecture specification. 

In summary, architecture conformance analysis techniques highlight the often neglected
topic of \emph{architecture} and offer a way to automatically analyse its impact on
the code. The technique is fragile in case of defective architecture specifications but
offers possibilities to actively discuss the influence of architectural decisions on the system implementation. We apply ConQAT for architecture conformance analysis.
A detailed description of its architecture conformance feature can be found in the work of \cite{Deissenboeck2010}.

\subsection{Quality Models and Quality Assessment Models}
\label{sec:quality-models}
Besides static analysis tools, we evaluate the benefit of quality models.
Following \cite{2011_iso_standard_25010}, quality models ``categorise product quality into characteristics.''
Hence, they mainly define what software quality consists of. In a broader sense, as discussed by
\cite{2009_deissenboeckf_purposes_scenarios}, quality models are able to assess or even predict the quality of a software system.
Hence, quality models exist, among others, as simple taxonomies, guidelines, checklists, metrics or stochastic models.
The aim of quality models is to make the abstract concept of \emph{software quality} more tangible. This can result in taxonomies used as checklists for quality
requirements, metrics used for an overall quality assessment or analyses for detecting critical parts of
a software system.

The work on software quality models began in the 1970s with early taxonomies from 
\cite{1978_boehmb_software_quality} and \cite{McCall.1977}. They broke quality down into 
what is colloquially called ``-ilities'' such as \emph{reliability} or \emph{maintainability}. 
This influenced the standard \cite{2003_iso_standard_9126_1} and its successor
\cite{2011_iso_standard_25010}. These taxonomies have shown to be too abstract
to be used by developers in their daily work~\citep{2009_wagners_quality_models_practice}. The
metrics proposed have proven to be difficult and the quality attributes are hard to be refined further  \citep{1996_kitchenhamb_quality_elusive_target,al-kilidar:isese05}. Therefore, the standards are not yet
widely used \citep{wagner:tum-i129}. Other researchers proposed more structured quality
models. For example, \cite{Dromey.1995} used a generic quality model split into quality
attributes, components and component properties as well as their interrelations. This way, he
could express the impacts of specific properties on the quality attributes more precisely. 

The German research project Quamoco\footnote{\url{http://www.quamoco.de/}} used the preliminary
work of its project partners~\citep{deissenb:icsm07,ploesch_et_al_2009} to develop a detailed and
explicit meta-model for quality models and build a broad, completely operationalised quality model
called \emph{base model}~\citep{wagner:icse12}. 
On the top level, the Quamoco quality models use quality characteristics well-known from \cite{2011_iso_standard_25010}, on the bottom level, they apply concrete
measures. 
In the base model, these are static measures collected either
automatically by ASA tools or manually by reviews. 
The base model is split into several \emph{modules}, a root module and several technology-specific ones.
Between quality characteristics
and low level measures, \cite{wagner:icse12} introduced the concept of a product factor which is similar
to the component properties of \cite{Dromey.1995}. Product factors bridge the gap between the different abstractions by being more directly 
measurable as well improving the traceability of the impact on the quality characteristics. 

Furthermore, \cite{wagner:icse12} describe a quality assessment method based on the aforementioned
quality model. It uses the collected data for the measures together with aggregation rules, so-called \emph{evaluation specifications}, for the product factors. 
Such specifications define how the measures associated to product entities translate
into the degree of presence of product factors. They are also calibrated using a large number of open-source
systems \citep{Lochmann2012}. 
There are similar specifications that describe (using weights) how the various product factors impact the quality
characteristics. The overall assessment result is then a value between 1 and 6 according to German school grades (1 is the best, 6 the worst). In empirical validations \citep{wagner:icse12}, such results have shown to be significantly correlated to expert opinions. The base model and tooling are available as open source, the base model is also accessible
in a Web version.\footnote{\url{http://www.in.tum.de/webportal/explorer.html}}

We introduced an example bug patterns above: the rule 
\emph{uncallable method defined in anonymous class} of FindBugs. It can measure the product factor \emph{Uselessness of a 
Method} which is part of the module on object orientation. Each product factor has a short description:
\begin{quote}
A method is useless if it is never referenced but nevertheless explicitly defined.
\footnotesize{\emph{Note:} This factor regards completely unnecessary code which could be deleted without any effect, whereas the
``unnecessarily complicated'' factor regards code that is too complicated due to an apparent programming mistake.}
\end{quote}
This is a factor that influences several quality characteristics of a software system negatively.
The quality model contains impacts to \emph{Resource Utilisation}, \emph{Analysability} and
\emph{Functional Correctness}. These, in turn, influence higher-level characteristics. For example,
\emph{Resource Utilisation} influences \emph{Performance Efficiency} which influences \emph{Quality}.
For evaluating the product factor and its influenced quality characteristics, we need to measure
the degree to which the product factor is present in a product. To do so, we have assigned measures
to the product factor. In our example, these are \emph{Uncallable method defined in anonymous class}
and \emph{Never called} for Java. Both can be collected automatically by the tool FindBugs. We collect
the number of findings in relation to the size of the product (e.g.~measured in LoC) and assign an evaluation result to the product
factor using an evaluation specification. Hence, many findings would lead to
a large degree of uselessness of methods which, in turn, could lead to a bad grade such as a 5. This way,
the measurements and evaluations at large constitute an overall grade of product quality.

To make building Quamoco quality models and using them in the assessment of software systems feasible, the 
project developed extensive tool support. There is a complete tool chain containing the Quamoco 
quality model editor for building the model and the assessment toolkit ConQAT for measuring and
calculating the assessment results~\citep{deissenboeck2011quamoco}. We used the Quamoco
base model together with the Quamoco tool chain in our study.

\subsection{Procedure}
\label{sec:proc}

This section explains the planned milestones of our investigation (Steps 1--4). It explains the start of our research (Step 1) and addresses our
research questions, i.e.~which data have to be collected and how to achieve that (Step 2) as well as how and under which conditions
our analyses have to be carried out (Steps 3--4). Steps 2 and 3 take place during a single, collaborative two-week
\textit{sprint} per participating enterprise with at least one sprint meeting at the 
beginning, the middle and the end of the sprint.
During a sprint, the respective study participant has to provide support for technical 
questions, such as check-out of the source code or configuration of the build process and 
has to attend the sprint meetings.

\subsubsection{Step 1: Workshops and Interviews}
\label{sec:proc_step1}
First, we conduct a series of workshops and interviews to motivate industrial partners
to participate in our project and then to understand their context and their needs. We start with an information event where we explain the
general theme of transferring QA techniques and propose first directions.
With all study participants (i.e.~the companies that agree to join the project), we conduct a kick-off meeting
and a workshop to create a common understanding, discuss organisational issues and plan the complete schedule. In addition, the
SPs present the corresponding study objects and their needs concerning software quality.
To intensify our knowledge of these systems and problems, for each participant, we perform a semi-structured interview with two
interviewers and a varying number of interviewees (i.e.~SP including optional staff). Both interviewers take notes and consolidate them. We then compare all interview
results to find commonalities and differences. After that, we hold one or two consolidation workshops to discuss our results and plan
the further investigations.

\subsubsection{Step 2: Raw Data Collection}
\label{sec:proc_step2}
The \textit{source code} of at least three versions of the study objects, e.g.~major releases chosen by the companies, is
retrieved for the application of the chosen techniques for RQ~1 to analyse effects over time. For bug pattern detection and
architecture conformance analyses, we retrieve or build \emph{executables} packed with debug symbols for each of these configurations. 
For architecture conformance, we also need an appropriate \emph{architecture specification}.
Accordingly, all study participants have to provide project artefacts as far as available, i.e.~source code, build environment and/or
debug builds, as well as documentation of source code, architectural
specification and project management information. Since the project budget
is limited, we cannot afford the additional effort of creating these project-specific artefacts if unavailable.

\subsubsection{Step 3: Measurement and Analysis}
\label{sec:proc_step3}
We apply each technique to the gathered raw data via corresponding tool runs and inspect the 
results, i.e.~findings and statistics. To get comparable results, we follow a generic procedure 
for each analysis technique which is presented here. A detailed description of the respective  
process for each technique can be found in the following paragraphs.

\begin{procedureList}
\begin{enumerate}
	\item \emph{Introduction} (once per study object):\newline
	Individual steps for preparation such as
	completion of technical and conceptual prerequisites,
	installation and configuration of tools,
	setup of aggregation mechanisms, filters and visualisations to 
	obtain tangible results.

	\item \emph{Application} (once per SO version or analysis run):
	\begin{enumerate}
		\item \emph{Readjustment} (optional): 
			Readjustment of the tools according to version specific 
			characteristics (e.g.~new path, excluded code, filtering of false 
			positives or reconfiguration of the tools).
		\item \emph{Run analysis:} 
			Run of the ASA tool or Quamoco tool chain, generation of results.
		\item \emph{Inspection:} 
			Inspection, filtering and discussion of findings, 
			identification of false positives.
	\end{enumerate}
\end{enumerate}
\end{procedureList}

To provide answers to RQ~1, we consider problems arising and efforts spent while following the 
structure of the presented procedure for each technique.
The tool runs as well as the application of the quality model and the discussion with the study 
participants in sprint meetings enable us to derive answers for RQ~2.1 (the classification of 
findings as defects, their discussion and analysis) and RQ~3.1 (the comparison of individual ASA 
results with the results of the quality model).

One person
per technique carries out the presented steps (I.1--II.3) for all SOs and versions. We discuss below
in detail how we perform the procedure for each of the investigated techniques.

\paragraph{Code Clone Detection (CD).}

We use the clone detection feature \citep{Juergens2009Clone} of ConQAT~2.7 for all SOs. In conventional clone detection, the configuration consists of two parameters: the minimal clone length and the source code path. In gapped clone detection, parameters such as maximal allowed number of gaps per clone and maximal relative size of a gap are required in addition. Based on the experience of our group and initial experimentation, we set the minimal clone length to 10 lines of code, the maximal allowed number of gaps per clone to 1 and the maximal relative size of a gap in our analysis to 30\%. After providing the needed parameters we run the analysis.

To inspect the analysis metrics and particular clones, ConQAT provides a list of clones, lists of instances of a clone, a view to compare files containing clone instances and a list of discrepancies for gapped clone analysis. This data is used to recommend corrective actions. In a series of clone detection runs over different SO versions, we monitor trends, i.e.~how the metrics evolve.

\paragraph{Bug Pattern Detection (BP).}
For Java-based systems we install and configure FindBugs 1.3.9 and PMD\footnote{\url{http://pmd.sourceforge.net}} 4.2.5. In C\#.NET contexts we use Gendarme\footnote{\url{http://www.mono-project.com/Gendarme}} 2.6.0  and FxCop 10.0. 

Aside from applying \emph{all rules}, we choose two additional tool settings which we consider to be relevant for the SOs to gain two focused quality perspectives:
\begin{enumerate}
  \item \textit{Selected categories} addressing correctness, performance and security.
  \item \textit{Selected rules} for unused or poorly partitioned code and bad referencing.
\end{enumerate}
The tool settings are determined during preliminary analysis test runs. Categories and rules which are considered not important -- based on discussion with the study participants as well as requirements non-critical to the SOs' application domains -- are ignored during rule selection.
For additional and language-independent metrics (e.g.~lines of code without comments; code-comment ratio; number of classes, methods and statements; depth of inheritance and nested blocks; comment quality) as well as for result preparation and visualisation, we apply ConQAT.
To simplify the issue of defect classification for our investigation, we only distinguish between rules for \textit{bugs} (obvious defects), 
\textit{smells} (simple to very complex heuristics for latent defects) 
and 
\textit{pedantry} (less critical issues with focal point on coding style). 
The readjustment of the tools to different versions involves a revision of the rule selection, a filtering of findings and an adjustment of the list of files to analyse.
Next, we analyse the finding reports resulting from the tool runs. This step involves besides the filtering of findings, primarily by rule criticality or finding frequency,  the inspection of source code to confirm the severity of and confidence in these findings and to determine corrective actions. To get feedback and to confirm our conclusions from the findings, we discuss them with our study participants during a workshop.

\paragraph{Architecture Conformance Analysis (AC).}
We start with the analysis of the architectural specification of each SO and extract relevant information.
Subsequently we contact system architects and validate our perception.
After installation of ConQAT, we create a reflexion model containing components and their mapping to code parts (e.g.~packages, name\-spaces, classes). We exclude code parts from the analysis that are not relevant (e.g.~certain external libraries).
Then ConQAT checks the conformance of the system to the reflexion model. Every existing dependency that 
is not allowed by the architectural rules represents a defect. Defects are visualised by the tool on the level of 
components and on the level of classes.
To eliminate tolerated architecture violations and to validate the created reflexion model, we discuss and classify every 
found defect together with the corresponding study participant.
This allows us to group similar defects and to gain a general understanding.

Architecture conformance analysis requires an architectural specification of the system which circumscribes allowed and forbidden
dependencies between logical components and their mapping onto the code. We are aware that this specification might be missing 
in some of our study objects. Unfortunately, project constraints (time, budget) hinder us and our study participants to 
reconstruct an architectural specification at least for our investigation if it is missing. 
Moreover, we want to emphasise that this reconstruction is associated with large efforts. 
This circumstance will also influence the evaluation and discussion of this technique.

\paragraph{Quality Model (QM).}
\label{sec:proc_quamod}
 
We make use of the existing Quamoco base model\footnote{\url{http://www.in.tum.de/webportal/explorer.html}} with its operationalisations for Java and C\# to assess
the quality of the SOs. We analyse all of them with respect to the base model using the Quamoco tool chain \citep{wagner:icse12}. 
Accordingly, we install Quamoco which includes a setup of ConQAT. 
To adapt Quamoco for our study we 
select the appropriate model module (Java or C\#) and configure the paths to the source code and 
executables. After running the analysis, we collect tool measurements, grades for the 
quality characteristics and their relations.

To answer RQ~3.1, we use 
the clone coverage value 
	from the clone detection results, 
the total number of findings per thousand lines of code and 
the number of critical findings 
	from the bug pattern results as well as 
the number of architecture violations 
	from the architecture conformance results where 
available. Using this information, we order the SOs by their rank regarding each of these sums. The higher the
number of findings or violations, the lower the rank. Similarly, we form a rank order using the overall quality
grade from the Quamoco assessment. This allows us to compare each ranking from the static analysis
tools directly with the aggregated result from the Quamoco assessment. 

\subsubsection{Step 4: Questionnaire}
\label{sec:StepFourQuestionnaire}

We use two kinds of questionnaires during the project. The first one aims at answering RQ 3.2, the second one is intended to  
address RQ 2.2. 

\paragraph{Comparison of Quality Model Results and Study Participants' Opinions.}
By means of a questionnaire, the companies estimate six characteristics of their systems' quality which are taken from the Quamoco quality model:
functional suitability, performance efficiency, reliability, security, maintainability and portability. Concerning 
maintainability, which is well elaborated in the Quamoco quality model, we are interested in four sub-attributes
which we include in the questionnaire: analysability, modifiability, reusability and testability.
The companies are allowed to give two estimations for each of the attributes: 
\begin{enumerate}
 \item An estimation for each attribute of the quality of their product, ranging in 7 levels from \textit{Insufficient} to 
 \textit{Excellent}.
 \item An estimation of how sure they are of their ratings, ranging in 7 levels from \textit{Unconfident} to \textit{Confident}.
\end{enumerate}
Finally, we compare the outcome of the questionnaires with the rating offered by the quality model.

\paragraph{Experience in Static Analysis Techniques.}
First, we evaluate the experience of the participating enterprises regarding software quality as well as static analysis
techniques. Second, we determine the perceived usefulness of ASA techniques for SMEs and whether our study participants plan
to use the presented techniques in their future projects. Thus, we perform a survey on our study subjects using a questionnaire
containing nine questions (Q1--9) which can be found in Tables~\ref{tab:rq2.2_1} and \ref{tab:rq2.2_2} in Appendix~\ref{sec:questionnairedetails}. 
The study participants fill out this questionnaire and
we evaluate the answers. To avoid the risk of biased or too narrowly formulated answers, we use open and closed questions.
This way we contribute to RQ~2.2.

\subsection{Protocol of Study Preparation (Step 1) and Execution (Steps 2--4)}
\label{sec:result:step1}

We held the information event of Step~1 of our procedure in July 2009 and invited more than thirty SMEs 12 of
which finally participated. Of these companies, five committed to act as study participants. The other companies 
were not showing commitment because of reasons we did not investigate further. 
We conducted the kick-off meeting in March 2010, the
interviews between March and July and, finally, 
two consolidation workshops in July 2010. We 
used these meetings to analyse the experience of the enterprises, to present and identify 
interesting topics related to software quality and to agree on a procedure and milestones for 
our collaboration.

Steps~2 and 3 were conducted in two-week sprints with each of the five study participants 
from October 2010 to January 2011. 
At the beginning of each sprint, the study 
participants gave us an introduction to their study object, its background and rationales. We 
met technical prerequisites (Step 2) such as remote access to the code repository and build
environment, the installation of required libraries and IDEs, the availability of architectural 
information and the identification of irrelevant parts of the study objects such as generated code.
Additionally, we explained to the study participants how we were going to perform the analyses of 
the study objects and which goals we wanted to achieve.

We conducted Step~3, the measurement and analysis concentrating on the three ASA techniques, 
during the whole sprint. Due to time constraints, quality 
models could not be considered in the sprints and were postponed to the time after the sprints.
After one week we held an intermediate meeting where we discussed potential problems with the 
study participants and presented the progress of our analyses. 
At the end of each sprint, we organised a 
final meeting with them to discuss of our findings and their 
criticality. Furthermore, we explained how the SPs could introduce and apply the
three techniques.
In January 2011 and after the sprints, we carried out the quality model analyses for each study object.
The analyses could benefit from the experience and knowledge we had gained in the sprints.

In addition, we distributed a questionnaire for RQ~3.2 (Tables~\ref{tab:rq2.2_2} \& \ref{tab:rq2.2_1}) to each of the 
study participants. 
We held the final project workshop at the end of February 2011. We presented the results and findings of the 
three ASA techniques and the quality model to all study participants and discussed the outcomes. 
After the final project workshop, we provided each of them with another questionnaire (Table~\ref{tab:rq3.2}) to answer RQ~2.2.

\section{Results}
\label{sec:results}

In the following, we present results for each technique for research questions RQ~1 and 2 and describe our findings for RQ~3 using a quality model.

\subsection{Code Clone Detection}
\label{sec:results_clones}

\subsubsection{RQ 1.1 -- Technical Problems}
Code clone detection turned out to be the most straightforward and least complicated of the three techniques. It has some technical limitations, however, that could hinder its application in certain software projects.

A major issue was the analysis of projects containing both markup and program code like JSP or ASP.NET. Since ConQAT supports either a markup language or a programming language during a single analysis run, the results for both languages need to be aggregated. To avoid this complication and concentrate on the code implementing the application logic, we only considered the code written in the programming language and ignored the markup code. Nevertheless, spending the effort of combining markup and program clone detection would provide more accurate results.

Another technical obstacle was filtering out generated code from the analysed code base. In one SO, large code portions were generated by the parser generator ANTLR%
\footnote{\url{http://www.antlr.org}}. We excluded such code files from our analysis using regular expressions.

Finally, the technique as applied is limited in the types of detectable clone classes. One may come across \emph{semantic clones}, i.e.\ code fragments which exhibit highly similar input-output behaviour but they differ too much to be recognised as normal or gapped clone instances. Chapter~\ref{sec:discussion} refers to possible solutions to these issues.

\subsubsection{RQ 1.2 -- Spent Effort}

\begin{table}[t]
\footnotesize
\begin{center}
\renewcommand{\arraystretch}{1.3}	
  \begin{tabular*}{\textwidth}{L{1.5cm}|L{2.08cm}|L{1.3cm}|L{1.6cm}|L{1.91cm}|@{\extracolsep{\fill}}L{1.3cm}}
  \textbf{Phase} & \textbf{Work Step} & \textbf{Clone\newline Detection} & \textbf{Bug Pattern Det.} & \textbf{Architecture\newline Conformance} & \textbf{Quality Model} \\\hline\hline
\multirow{2}{2cm}{I.\newline Introduction} & \multirow{2}{2cm}[-2ex]{Individual procedure for preparation}
                            
& \multirow{2}{2cm}{$\leq 0.5h$} 
& setup: \newline $\leq 1h$ 
& conception: $\leq 1d$ 
& \multirow{2}{2cm}{$\leq 0.5h$} \\					 
& 
&  
& aggregation: $\leq 0.5d$ 
& reflexion model: $\leq 1h$ & 
\\\hline
\multirow{3}{2cm}{II.\newline Application}  & 
1. Readjustment &  $\leq 0.5h$						& $\leq 0.5h$		&  $\leq 0.5h$	& n/a 
\\\cline{2-6} & 
2. Run analysis		& $\leq 5min$ 						& $\leq 1h$  		& $\leq 10sec$	& $\leq 15min$ 
\\\cline{2-6} & 
3. Inspection		& $\leq 1h$, gapped: $\leq 1.5h$	& $\leq 0.5h$	& $\leq 0.5h$ & $\leq 1h$
 \\\hline
  \end{tabular*} 
  \caption{Maximum effort spent (RQ 1.2) across all study object versions for applying each of the techniques, following the presented steps in \ref{sec:proc}. 
  \textbf{Legend:} h \dots hour(s), d \dots day(s)
  \label{tab:efforts}}
\end{center}
\end{table}

Table~\ref{tab:efforts} gives an overview of the needed effort.
The time required to introduce clone detection is small compared to the other two ASA
techniques under study. The ease of introduction of clone detection is achieved due to the minimalist configuration of the analysis which, in the simplest case, includes the path to the source code and the minimal length of a clone. 
Our estimations were made under the assumption that relevant tool manuals \citep{DeissenboeckFeilkasEtAl2010:ConQATBook} 
have already been consulted.

For all SOs, it took less than an hour to configure clone detection, to get the first results and to investigate the longest and most frequent clones. Running the analysis itself took less then five minutes.

In case of gapped clone detection, it took a considerable amount of time to determine whether a discrepancy is intended or whether it represents a defect. To speed up the rest of our procedure, ConQAT supports the intended discrepancies to be fingerprinted and excluded from further analysis runs.

\subsubsection{RQ 2.1 -- Found Defects} 
\label{sec:results_clones:rq21}

\begin{table}[t!]
\begin{center}
  \begin{tabular}[\textwidth]{c|c|R{1.4cm}|R{1.2cm}|R{.9cm}|R{1.4cm}|R{1.2cm}|R{1.5cm}}
    \textbf{SO} & \textbf{Ver.} & \textbf{Analysed Units [kUnits]} & \textbf{Cloned Units [kUnits]} &\textbf{Blow-up\newline [\%]} & \textbf{Unit\newline Coverage [\%]} & \textbf{Longest Clone [Units]} & \textbf{Most Clone Instances} \\\noalign{\hrule height 1pt}
      & I   &  15.9 &  3.5 & 119.5 & 22.2 & 112 &  39 \\\cline{2-8}  
    1 & II  &  25.3 &  5.8 & 118.9 & 23.0 & 117 &  39 \\\cline{2-8}    
      & III &  32.3 &  7.8 & 119.2 & 24.0 & 117 &  39 \\\noalign{\hrule height 1pt}
      & I   &  35.4 & 14.3 & 143.1 & 40.5 &  63 &  64 \\\cline{2-8}  
    2 & II  &  41.6 & 18.9 & 150.2 & 45.4 & 132 &  47 \\\cline{2-8}    
      & III &  39.9 & 14.6 & 137.4 & 36.7 &  89 &  44 \\\noalign{\hrule height 1pt}
      & I   &  51.7 &  9.4 & 114.5 & 18.2 &  79 &  21 \\\cline{2-8}  
    3 & II  &  56.8 &  8.6 & 111.2 & 15.1 &  52 &  20 \\\cline{2-8}    
      & III &  61.6 &  8.4 & 110.0 & 13.7 &  52 &  19 \\\noalign{\hrule height 1pt}
      & I   &   8.9 &  6.0 & 238.8 & 68.0 & 217 &  22 \\\cline{2-8}  
    4 & II  &  22.4 & 17.3 & 309.6 & 77.6 & 438 &  61 \\\cline{2-8}    
      & III &  38.3 & 30.4 & 336.0 & 79.4 & 957 & 183 \\\noalign{\hrule height 1pt}
      & I   & 196.3 & 48.7 & 122.3 & 24.8 & 141 &  72 \\\cline{2-8}  
    5 & II  & 211.3 & 53.4 & 122.7 & 25.3 & 158 &  72 \\\cline{2-8}    
      & III & 208.6 & 53.2 & 122.8 & 25.5 & 156 &  72 \\\noalign{\hrule height 1pt}
  \end{tabular} 
  \caption{Results of code clone detection. 
  \textit{Units}: uncommented and normalised source code statement; \textit{kUnits}: 1000 Units.
   \label{fig:ucstudyobjects}}
\end{center}
\vspace{-10pt}
\end{table}

The Tables~\ref{fig:ucstudyobjects} and \ref{fig:gappedclones} show detailed results of conventional and gapped code clone detection for three versions of each SO. As explained in  Section~\ref{sec:methods_clones}, the column ``Analysed Units'' shows numbers smaller than the actual code size given in Table~\ref{fig:studyobjects}. 

The results of conventional clone detection can be interpreted as an indicator of bad design and software maintainability problems, but they do not point at actual defects. Nevertheless, these results indicate code parts as candidates for improvement. The following three types of clones 
were detected in all analysed systems to a certain extent: cloning of exception handling code, cloning of logging code and cloning of interface implementation by different classes.

In the analysed systems, unit coverage varied between 14 and 79\% (Table~\ref{fig:ucstudyobjects}). \cite{Koschke2007} reports on several case studies with unit coverage values between 7 and 23\% and one case study with a value of 59\%, which he defines as extreme. Therefore, SO~1, 3 and 5 exhibit normal clone rates according to Koschke. The clone rate in SO~2 is higher than the rates reported by Koschke, and for SO~4 it is extreme. Regarding maintenance efforts, the calculated blow-up for each system may indicate a risk. For example, version~III of SO~4 is more than three times bigger as its hypothetically equivalent system containing no clones. SO~4 shows that cloning can increase over time, while SO~3 reveals that it is possible to reduce the amount of its code clones.

\begin{table}[t!]
\begin{center}
  \begin{tabular}[\textwidth]{c|c|R{1.4cm}|R{1.2cm}|R{.9cm}|R{1.4cm}|R{1.2cm}|R{1.5cm}}
    \textbf{SO} & \textbf{Ver.} & \textbf{Analysed Units [kUnits]} & \textbf{Cloned Units [kUnits]} &\textbf{Blow-up\newline [\%]} & \textbf{Unit\newline Coverage [\%]} & \textbf{Longest Clone [Units]} & \textbf{Most Clone Instances} \\\noalign{\hrule height 1pt}
      & I   &  13.3 &  3.0 & 119.9 & 22.3 &  34 &  39 \\\cline{2-8}  
    1 & II  &  21.0 &  4.5 & 117.9 & 21.5 &  37 &  52 \\\cline{2-8}    
      & III &  27.1 &  6.0 & 117.4 & 22.1 &  52 &  52 \\\noalign{\hrule height 1pt}
      & I   &  24.3 &  4.6 & 116.3 & 19.0 & 156 &  37 \\\cline{2-8}  
    2 & II  &  34.7 &  8.7 & 123.2 & 25.0 & 156 &  37 \\\cline{2-8}    
      & III &  37.1 &  9.4 & 123.7 & 25.3 & 156 &  37 \\\noalign{\hrule height 1pt}
      & I   &  46.7 & 12.0 & 124.4 & 18.2 &  73 & 123 \\\cline{2-8}  
    3 & II  &  46.1 & 10.0 & 120.0 & 15.1 &  55 &  67 \\\cline{2-8}    
      & III &  49.1 & 10.0 & 118.6 & 20.5 &  55 &  64 \\\noalign{\hrule height 1pt}
      & I   &   7.8 &  4.5 & 192.1 & 58.6 &  42 &  34 \\\cline{2-8}  
    4 & II  &  18.8 & 11.0 & 206.2 & 59.8 &  51 &  70 \\\cline{2-8}    
      & III &  32.2 & 19.2 & 211.1 & 59.5 &  80 & 183 \\\noalign{\hrule height 1pt}
      & I   & 142.3 & 29.4 & 117.4 & 20.7 &  66 &  68 \\\cline{2-8}  
    5 & II  & 154.0 & 32.8 & 118.0 & 21.3 &  85 &  78 \\\cline{2-8}    
      & III & 151.9 & 32.7 & 118.2 & 21.5 &  85 &  70 \\\noalign{\hrule height 1pt}
  \end{tabular} 
  \caption{Results of gapped code clone detection.
  \textit{Units}: uncommented and normalised source code statement; \textit{kUnits}: 1000 Units.
  \label{fig:gappedclones}}
\end{center}
\vspace{-10pt}
\end{table} 

Code clones are considered harmful because they increase the chance of unconscious, inconsistent changes, which can lead to faults~\citep{Juergens2009}. These changes can be detected using gapped clone detection. Table~\ref{fig:gappedclones} shows corresponding results. We found a number of such changes in the cloned code fragments, but we could not classify them as defects, because we lacked knowledge about the SOs. Despite our workshop discussions, the study participants were not able to definitely classify these discrepancies as defects. This indicates that gapped clone detection is a more resource demanding type of analysis. Nevertheless, in some clone instances we identified additional instructions or deviating conditional statements compared to other instances of the same clone class. Gapped clone detection does not cross method boundaries, since experiments showed that inconsistent clones that cross method boundaries in many cases did not capture semantically meaningful concepts~\citep{Juergens2009}. This explains why metrics such as cloned units or unit coverage (Table~\ref{fig:gappedclones}) may differ from values observed with conventional clone detection (Table~\ref{fig:ucstudyobjects}). The smaller numbers in column ``Analysed Units'' of Table~\ref{fig:gappedclones} result from the exclusion of units not meaningful for the gapped variant, e.g.\ source code outside from method definitions.

\subsubsection{RQ 2.2 -- Perceived Usefulness} 

Following the feedback obtained from the questionnaire (see Q1--Q8 in Tables~\ref{tab:rq2.2_1} and \ref{tab:rq2.2_2} in Appendix~\ref{sec:questionnairedetails}), two study participants had limited experience with clone detection, the other three did not consider it at all (Q2). Three participants estimated the relevance of our clone detection results to their projects as high, the other two estimated it as medium relevant (Q3). 
SO~2 and 4 had high clone rates. The participant responsible for SO~2 considered this as
medium relevant. For SO~4, the SP considered its clone rates as highly relevant. 
The relevance is underpinned by one SP's argument that ``clone detection is only feasible with tool support'' which we demonstrated (see Q1 in Table~\ref{tab:rq2.2_2}). Another interesting statement was that ``clones are necessary within short development cycles.'' Finally, all SPs evaluated the importance of using clone detection in their projects as medium to high and plan to introduce this technique in the future (Q5).

\subsection{Bug Pattern Detection}
\label{sec:results_patterns}

\subsubsection{RQ 1.1 -- Technical Problems}
\label{sec:results_patterns:rq11}
We can confirm that bug patterns are a powerful technique to gather a vast variety of information about potentially defective code. However, most of its effectiveness and efficiency is achieved through carefully done, project-specific fine-tuning of the many setscrews available. This is confirmed by \cite{Boogerd2009} and \cite{Ruthruff2008}.
In the following, we mention three important issues:

First, the impact of findings on product quality factors or characteristics of interest and their consequences for the project (e.g.~corrective actions, avoidance or tolerance) were difficult to determine by the tool-provided rule categories, the severity and confidence information. Based on our experience, we identified the following study object characteristics this impact depends on:
\begin{itemize}
  \item Required usage-level qualities, e.g.~security, safety, performance, usability 
  \item Required internal qualities, e.g.~code maintainability, reusability 
  \item Technologies, i.e.~language, framework, platform, architectural style
  \item Criticality of the context, the findings belong to, e.g.~platform or driver code 
\end{itemize}

Second, some rules exhibited many false positives, either because their technical way of detection is fuzzy or because a definitely precise finding is considered not relevant in a project-specific context. The latter case requires an in-depth understanding of each of the rules, the impacts of findings and, subsequently, a proper redlining of rules as pedantry or, actually, irrelevant. We neither measured the rates of false positives nor investigated costs and benefits thereof as our focus lay on the identification of the most important findings only. 

Third, due to restricted selection and filtering mechanisms within the tools as well as lack of knowledge about the SO life cycles, we were hindered to apply and calibrate appropriate rule selectors and findings filters. We saw that the usefulness of results is crucially influenced by the conversion of project-specific information on rule impacts into queries for rule selection and findings filtering.
The tools largely differ in their abilities to achieve this task via their graphical or command-line interfaces. 

We addressed the first two issues by group discussion also with our study participants and improved rule selection and findings filtering to principally avoid the finding reports to get overloaded or prone to false positives of the second kind. The third issue could also only be compensated by manual efforts. As most finding reports were quite homogeneously encoded and technically well accessible, we utilised ConQAT to gain statistical information for higher-level quality metrics as listed in Step 3 of our procedure. Chapter~\ref{sec:discussion} refers to approaches to better overcome these issues.

\subsubsection{RQ 1.2 -- Spent Effort}
Table~\ref{tab:efforts} gives an overview of efforts.
We achieved the initial setup of a single bug pattern tool in less than an hour. This estimation excluded the time needed to gain previous knowledge about the internal structure of the SO such as, e.g.~its directory structure and third party code.
We used the ConQAT framework to flexibly run the tools in a specific setting 
and for further processing of the finding reports. Having good knowledge of this framework, we completed the analysis setup for an SO (selection of rules, adjustment of bug pattern parameters and framework setup) in about half a day.

The runs took between a minute and an hour depending on code size, rules selection and other parameters. Hence, bug pattern detection should at least be selectively included into automated build tasks. Part of the rules are computationally complex and some tools frequently required more than a gigabyte of memory. 
The manual effort after the runs can be split into inspection and readjustment.
The inspection of a report took us a few minutes up to half an hour.
Readjustment of the rule selector and the findings filter requires deep knowledge of the type, objectives and evolution history of an SO. As we could not gain this knowledge in our two-week sprints (see Section~\ref{sec:proc}), readjustment was only done roughly and, hence, took no more than half an hour.

\subsubsection{RQ 2.1 -- Found Defects}
\label{sec:results_bugpatterns:rq21}
We conducted bug pattern analysis in three selective tool settings according to Step~3 but only for one version of each SO. Table~\ref{fig:bugpatternstats} summarises noticeable findings that have been most critically rated by the tools, exhibited relatively high frequencies or have been extraordinarily remarkable.
For all SOs, the filtered finding reports confirmed the defects focused or expected by these settings. We used ``*'' to label findings which have been explained to our study participants and consensually confirmed as critical at the final project workshop. 
Without going into the quantities and details of single findings, we summarise language-specific results:

\begin{description}
\item[\textbf{C\#}] Among the rules with the most critical or frequent findings in SO~1 and 2, FxCop and Gendarme reported \textit{empty exception handlers}, \textit{visible constants} and \textit{poorly structured code}. There was only one kind of findings related to correctness consensually considered to be critical in SO~1, namely unacceptable loss of precision through \emph{wrong cast during an integer division} used for accounting calculations (Table~\ref{fig:bugpatternstats}).

\item[\textbf{Java}] Among the rules with the most critical or frequent findings in SO~3, 4 and 5, FindBugs and PMD reported \textit{unused local variables}, \textit{missing validation of return values}, \textit{wrong use of serialisable} (mainly SO~3) and extensive \textit{cyclomatic complexity}, \textit{class or method size}, \textit{nested block depth}, or \textit{parameter list} (SO~3, 4 and 5). There have only been two kinds of findings related to correctness consensually considered to be critical, both in SO 5, namely foreseeable \emph{access of a null pointer} and an \emph{integer shift beyond 32~bits} in a basic date-time calculation component (Table~\ref{fig:bugpatternstats}).
\end{description}
Independent of the programming language and concerning security and  stability, 
we detected the pattern \textit{problematic method call} in four out of five SOs (e.g.~frequently in SO 3 and 5: \textit{constructor calls overwritable method}) and found a number of defects related to \textit{error prone handling of pointers}.
Concerning maintainability, the SOs exhibited \textit{missing or unspecific handling of exceptions}, manifold \textit{violations of code complexity metrics} and various forms of \textit{unused code}.

\begin{table}[t!]
\begin{center}
\setlength{\tabcolsep}{3pt}
\renewcommand{\arraystretch}{1.2}
\begin{tabular*}{\textwidth}{p{1.2cm}@{\extracolsep{\fill}}|p{4.6cm}|c|c|c|c|c|p{2.2cm}}
\multirow{3}{1.2cm}{\textbf{Tool} (Lang.)} & \multirow{3}{4.9cm}{\textbf{Rule or (G)roup of Rules}} & \multicolumn{5}{c|}{\textbf{Study Objects (Ver.)}} & \multirow{3}{2.5cm}{\textbf{Most affected Qualities}} \\
& & 
$\;\;1\;\;$ &    
$\;\;2\;\;$ &    
$\;\;3\;\;$ &    
$\;\;4\;\;$ &    	
$\;\;5\;\;$ &    
\\
& & III 
& III 
& III 
& II & II & \\
& & C\# & C\# & J & J & J &
\\\hline\hline
\multirow{2}{1.3cm}{FxCop (C\#)} & 
	Empty/general except.\ handlers 
	& 106 & 47 & - & - & - & \multirow{3}{2cm}{Maintainability} \\
& 
	Nested use of generic types   
	& 17 & 24 
	& - & - & - & \\\cline{1-7}
\multirow{2}{1.3cm}{Gend\-arme (C\#)} & 
	Deep namespaces with depth $> 4$
	& 0  & 35 & - & - & - & \\\cline{8-8}
	& Visible constants 
	& 338 & 18 & - & - & - & Security \\
    & Suspicious type conversion
	& 3* & 0 &	- &	- & - & Correctness \\\hline
\multirow{2}{1.3cm}{Gend., PMD} & 
	G: Problematic method calls
	& 2 
	& 8 
	& i 
	& 0  & i 
	& Secur., stability \\\cline{8-8}
 	& G: Extensive class/method size or parameter count; too many fields
	& 20 
	& 1 & i & i  & i  & \multirow{4}{2cm}{Maintainability} \\\cline{1-7}
\multirow{2}{1.3cm}{PMD (Java)} & G: Empty methods
	&  -  &  -  &  i & 0 &  i & \\
    & Cyclomatic complexity $> 10$
	&  -  &  -  &  256 & 49 & 938 & \\\cline{1-7}
\multirow{4}{1cm}{Find\-Bugs (Java)} & 
	G: Unused fields/variables				
	&  -  &	- & 132 
	&  0  &  2  	& \\\cline{8-8}
 	& G: Inefficient string manipulation
	&  -  &	- & 49 
	&  0  & 0  & Performance \\\cline{8-8}
 	& G: Corrupted serialisable
	&  -  &	- & 12  
	&  0  & 0  & \multirow{2}{2cm}{Correctness} \\
    & Integer shift beyond 32 bits 
	&  -  &  -  &  0  &  0  &  4*  & \\\cline{8-8}
	& Return values not validated
	&  - &  -  & 32 
	&  0  & 0  & Correctn., secur.\\ 
    & Access of a null pointer
	&  -  & -  & 1   &  0  &  2*  & Secur., stability \\\hline
\multicolumn{8}{l}{\textbf{Maximum Metrics} (suggestions in parentheses)}\\\hline
PMD
    & Max.~cyclomatic complexity ($\le 10$)
	&  -  &  -  
	&  78 
	& 156 
  	& 216
  	& \multirow{2}{2cm}{Maintainability} \\\cline{1-7}
ConQAT & Max.~nested block depth ($\le 5$)		
	&  11 & 13 &  19 & 17 
				   	   & 14 & \\\hline

  \end{tabular*} 
  \caption{Summary of noticeable bug pattern findings and maximum metrics. \textbf{Legend:} Cells contain the number of findings or a maximum value, 
  - \dots not applicable, 
  i \dots noticeable, but PMD did not offer an appropriate way to exactly count the many findings; 
  * \dots regarded as consensually critical after group discussion; J \dots Java. 
  \label{fig:bugpatternstats}}
\end{center}
\vspace{-10pt}
\end{table}

\subsubsection{RQ 2.2 -- Perceived Usefulness}
\label{sec:results_bugpatterns:rq22}

According to the questionnaire (see Q1--Q8 in Tables~\ref{tab:rq2.2_1} and \ref{tab:rq2.2_2} in Appendix~\ref{sec:questionnairedetails}), all of the study participants considered our bug pattern findings to be medium to highly relevant for their projects (Q3).
The sample findings, we presented during our final project workshop, 
particularly Table~\ref{fig:bugpatternstats}, were perceived as critical enough to be treated if they had been found during the development of the SOs. However, the SPs did not perceive these findings as threats to the business success of the SOs.
The low number of consensually critical findings correlated well with the fact that the technique was known to all SPs and that most of them have good knowledge thereof and regularly used such tools in their projects, at least monthly, at milestone or release dates (Q1-2). 
Monthly use of FindBugs and PMD, as confirmed by the SP of SO~4, largely explains its relatively positive BP situation.
However, three of the SPs could gain additional education in this technique (Q4).
One stated that the presented analyses are ``often not feasible in projects externally conducted at the customer's site'' (Q8).
Nevertheless, all the participants indicated to use bug patterns as an important QA instrument in their future projects (Q5).

\subsection{Architecture Conformance Analysis}
\label{sec:results_archconf}

\subsubsection{RQ 1.1 -- Technical Problems}
We observe two general problems that prevent or complicate each architectural analysis: The absence of system architecture specification and the usage of \emph{dynamic patterns} of architecture design, i.e.\ a programming technique that increases flexibility through postponing type binding and identification until run-time.
 
For two of the SOs, no architectural specification was existent. In case of SO~3, the SP was aware of the lack of such documentation. Nevertheless, they feared that the time involved and the sheer volume of code to be covered exceeds the benefits. The need to update the specification, within several months or each time a new release is coming out, was stated as an additional argument.
For SO~5, the specification was missing because the project was taken over by SS~5 from a different organisation that was not documenting the architecture at all, for reasons we could not determine. The SP argued that any later specification of the architecture would be too expensive.

In SO~2, a dynamic design pattern was implemented 
so that no static dependencies could be found between defined components. The components belonging to the system are connected at run-time. Hence, architecture conformance analysis could not be applied. Additional configuration files used for dynamic patterns have not been taken into account by the tools we used.

Architecture conformance analysis needs two ingredients apart from the architecture specification: The source code and the executables of a system to resolve symbolic references. This is a potential barrier because the source has to be compilable to be analysed.

Another technical problem occurred while using ConQAT. Dependencies to components solely existing as executables were not recognised by the tool. Hence, all rules belonging to compiled components with missing source could not be analysed.

Finally, we could apply this technique to two systems without any technical problems. An overview of all SOs with respect to their architectural properties can be found in Table~\ref{fig:archstudyobjects}.

\subsubsection{RQ 1.2 -- Spent Effort}

Table~\ref{tab:efforts} gives an overview of efforts.
The first step in applying this technique would be the creation of an appropriate 
architecture specification. As explained in Section~\ref{sec:proc_step3}, we had to leave this out. Hence, we faced two situations: Either we could use the specification to build a reflexion model of the system or the study object was lacking such a 
specification.

In case of a missing specification, we asked the study participants to give reasons for the situation and why neither an a-priori specification nor an a-posteriori reconstruction was considered. For SO~3, the study participant argued that the cost-value ratio 
of a later reconstruction would be too bad considering the effort to maintain it. As the system is changing 
continuously and its 
development is distributed to many locations, the study participant decided against an 
architecture specification.
During the evolution of SO~5, such a specification was never produced. As SO~5 grew to a size of 560~kLoC 
and its inner structure got more complex, the effort for the creation of an architecture specification was 
considered too costly by its SP.
As explained in the last section, 
technical obstacles were hindering the application of the technique for SO~2.
The effort that was required to create the two other architecture specifications (SO~1, SO~4) could not be precisely estimated by the study participants. We think that it is difficult to exactly determine the whole time that was spent on creating a specification because this conceptual artefact is usually influenced by many project specific 
characteristics and tasks.

The most time consuming task was to inspect the architecture specification and to 
understand their content, to discuss them with system architects and to identify relevant code fragments, which took 
us in total up to one day.
The initial configuration of ConQAT including the creation of the reflexion model
could be achieved in less than one hour.

Table~\ref{fig:archstudyobjects} shows the number of modelled components and the rules that 
were needed to describe allowed connections. The analysis run finished in less than ten seconds. The time needed 
for the interpretation of the analysis results is of course dependent on the amount of defects found. For each defect, we were able 
to find the causative code parts within one minute. We expect that the effort needed for bigger systems will only increase linearly 
but will stay small in comparison to the benefit that can be achieved using architecture conformance analysis.

\begin{table}[t!]
\begin{center}
  \begin{tabular}[\textwidth]{c|l|c|C{3.5cm}|C{2.5cm}}
    \textbf{SO} & \textbf{Architecture} & \textbf{Version} & \textbf{Violating Component Relationships} & \textbf{Violating Class Relationships}  \\ \hline\hline
      & \multirow{3}{0.2\textwidth}{12 Components\newline 20 Rules} & I & 1 & 5 \\\cline{3-5}
    1 &  & II & 3 & 9 \\\cline{3-5}
      & & III & 2 & 8 \\\hline
    2 & dynamic pattern & n/a & n/a & n/a \\\hline
    3 & undocumented & n/a & n/a & n/a \\\hline
      &  \multirow{3}{0.2\textwidth}{14 Components\newline 9 Rules}& I & 0 & 0 \\\cline{3-5}
    4 & & II & 1 & 1 \\\cline{3-5}
      & & III & 2 & 4 \\\hline
    5 & undocumented & n/a & n/a & n/a \\\hline
  \end{tabular} 
  \caption{Architectural characteristics of the study objects (SO) 
  	\label{fig:archstudyobjects}}
\end{center}
\vspace{-10pt}
\end{table} 

\subsubsection{RQ 2.1 -- Found Defects}
\label{sec:results_archconf:rq21}
As shown in Table~\ref{fig:archstudyobjects}, we observed several discrepancies in the analysed SOs across nearly all 
versions. Only one version did not contain architecture violations. 
Overall, we found three types of defects in the analysed SOs. Each defect represents a code location showing a 
discrepancy to the architecture specification. All defects we found could be 
avoided if this technique had been applied. 
In the following, we explain the types of defects found and classified together with the study participants. 
All findings were rated by the study participants as critical.
 
\begin{itemize}
\item  \textit{Circumvention of abstraction layers:} Abstraction layers (e.g.\ presentation layer) provide a common 
way to structure a system into logical parts. The defined layers are hierarchically depending on each other, 
reducing the complexity in each layer and allowing to benefit from structural properties like exchangeability or 
flexible deployment of each layer. These benefits vanish once the layered concept is harmed by dependencies between 
layers that are not connected to each other. In our case, e.g.\ the use of the data layer by the presentation layer 
was a typical defect.

\item \textit{Circular dependencies:} We found undocumented circular dependencies between two components. We 
consider them -- whether or not documented -- as defects themselves because they affect principles of well designed architectures. Two components that are depending on each other can only be used together and can thus be 
considered as one component. The reuse of these 
components is strongly restricted. Their source code is harder to understand and to maintain. 

\item \textit{Undocumented use of common functionality:} Every system's internals make use of a set of common 
functions (e.g.\ for date-time manipulation) which are often grouped into a ``library'' component to be easily 
accessed and maintained. Thus, it is important to have an overview of where these functions are actually used. Our 
investigation showed such dependencies that were not covered by the architecture specification.

\item \textit{Data dependencies between components:} Aside from using external methods or remote procedures, components are often also using data structures defined in other 
components, e.g.\ classes, enumerations or user interface elements. Such dependencies can occur in various contexts, e.g.\ 
field declarations, method arguments or inheritances. Data dependencies between components are often not 
obvious and should therefore be documented.

\end{itemize}

\subsubsection{RQ 2.2 -- Perceived Usefulness}
Following the feedback gained from the questionnaire (see Q1--Q8 in Tables~\ref{tab:rq2.2_1} and \ref{tab:rq2.2_2} 
in Appendix~\ref{sec:questionnairedetails}), we observed that four out of five study participants did not know about the 
possibility of automated architecture conformance analysis (Q1). Only one of them already checked the architecture 
of their system, but in a manual way and infrequently. Confronted with the results of the analysis, all participants 
rated the relevance of the presented technique as medium to highly relevant (Q3). One of them stated that, as a new 
project member, it is easier to become acquainted with a software system if its architecture conforms to its 
documented specification (Q3). All participants agreed on the usefulness of this technique and plan its future 
application in their projects (Q5).

\subsection{Quality Models}

\subsubsection{RQ 3.1 -- Matching with ASA Results}

For this research question, we compare 
the evaluations of the SOs that one can get by manually looking at the numbers of findings and violations 
directly 
with the aggregated and weighted evaluation of the Quamoco quality model. 
This comparison gives us the opportunity to investigate whether 
the aggregation has an effect. 
We can find out some differences but not determine any exact causes for them.

We used the findings and violations from ASA as well as the Quamoco assessment grades
to form rank orders for the SOs. The result is shown in Table~\ref{tab:rq3.1pkg}. The Quamoco
model assessment ranks SO~2 as best with a grade of 1.5 and SO~5 last with a grade of 4.5.
There is a large spread considering the grade range from 1 to 6 and only the grades of SO~1 and
SO~3 are so close that the order might be questionable.
More details can be found in Tables~\ref{tab:rq3.1} and \ref{tab:quamocoRes} in Appendix~\ref{sec:questionnairedetails}.

\begin{table}[htpb]
  \begin{center}
  \renewcommand{\arraystretch}{1.1}

  \begin{tabular}{c|c|c|c|c||c|}
  &
  \multicolumn{5}{c|}
  {\textbf{SOs ranked by}} \\
  \rb{\textbf{Rank}}&
  \rb{\parbox{2.25cm}{\textbf{Clone Detection} \newline Clone Coverage}} &
  \rb{\parbox{2.25cm}{\textbf{Bug Pattern} \newline Total Findings per 1000 LoC}} &
  \rb{\parbox{2.25cm}{\textbf{Bug Pattern} \newline Critical Findings}} &
  \rb{\parbox{2.25cm}{\textbf{Architecture Conformance}\newline Violating Classes}} &
  \rb{\parbox{2.25cm}{\textbf{Quality Model} Overall Grade}}
  \\\hline\hline
 
  \textbf{1.} & \textbf{3} (13.7\%) & \textbf{4} (0.49)	& \textbf{2, 3, 4} (0)   & \textbf{4} (4) & \textbf{2} (1.5)  \\\hline
  \textbf{2.} & \textbf{1} (24.0\%) & \textbf{2} (0.67)	& 						 & \textbf{1} (9) & \textbf{4} (2.0)  \\\hline
  \textbf{3.} & \textbf{5} (25.5\%) & \textbf{5} (1.69)	& 						 & 				  & \textbf{3} (2.5)  \\\hline
  \textbf{4.} & \textbf{2} (36.7\%) & \textbf{3} (2.41) & \textbf{1} (3)		 & 				  & \textbf{1} (2.8)  \\\hline
  \textbf{5.} & \textbf{4} (79.4\%) & \textbf{1} (4.68)	& \textbf{5} (6)		 & 				  & \textbf{5} (4.5)  \\\hline
  \end{tabular} 
  \caption{Ranking (1. to 5.) of the study objects on the basis of ASA results and based on the grades given by the 
  quality model. \textit{Legend}: In brackets, clone coverage, number of findings or the grade given
  	by the quality model. 
	\label{tab:rq3.1pkg}
}
  \end{center}
\end{table}

\paragraph{Clone Coverage vs.\ Overall Grade.}
Extreme clone rates of more than 79\% lead to the last place of SO~4 while SO~3 ranks best with a rather low clone
coverage of almost 14\%. SO~1 and SO~5 have similar clone coverage around 25\%. SO~2 ranks fourth with a
clone coverage of almost 37\%. 
Surprisingly, the clone coverage results are not reflected in the quality grades. The worst ranked SO~4 in terms of
cloning reaches rank 2 in the quality grades. Instead, other study objects, which possess normal clone rates (SO~1, 
SO~5), are ranked worst by the quality model. In summary, code clone coverage as a single factor gives strongly
different quality rankings in comparison to the results given by the quality model.

\paragraph{Total Bug Pattern Findings vs.\ Overall Grade.}
The numbers of total bug pattern findings per 1000 lines of code range from 0.49 to 4.68. The highest rank is achieved by SO~4 with SO~2
as a close second. SO~3 and SO~1 are ranked as the worst. This ranking is close to the one from the quality
model assessment. The only difference is that SO~5 is on the third position when purely counting findings while the aggregated assessment ranks it last. Aggregation and weighting based on a calibration
over various open source systems introduces a view differing from purely counting findings.

\paragraph{Critical Bug Pattern Findings vs.\ Overall Grade.}
Because there were only a few critical bug pattern findings, we cannot completely separate them by ranks.
SO~2, SO~3 and SO~4 did not show any critical findings and are ranked best. SO~1 is on position 4 with
three critical findings and SO~5 on position 5 with six critical findings. Apart from the differentiation of three SOs,
this is an even better fit to the quality model ranking. The worst SOs are ranked exactly the same. Therefore,
concentrating on critical findings seems to provide a picture similar to the aggregated quality model result.

\paragraph{Architecture Conformance Violations vs.\ Overall Grade.}
Finally, the results for architecture conformance violations are limited because we only have two SOs for this
technique. We can rank them. SO~4 is on the best position with four violations while SO~1 ranks second
with nine violations. We cannot say if the overall ranking would be similar to the ranking from the quality model,
but the order of the two SOs is the same. The quality model also sees SO~4 as having a better quality than SO~1.
Hence, these evaluations seems similar.

\subsubsection{RQ 3.2 -- Comparison to the Study Participants' Opinions}

\begin{figure}[tpb]
  \centering
  \subfigure[Study object 1]{
    \label{fig:kiviatRQ32SO1}
    \includegraphics[height=4cm]{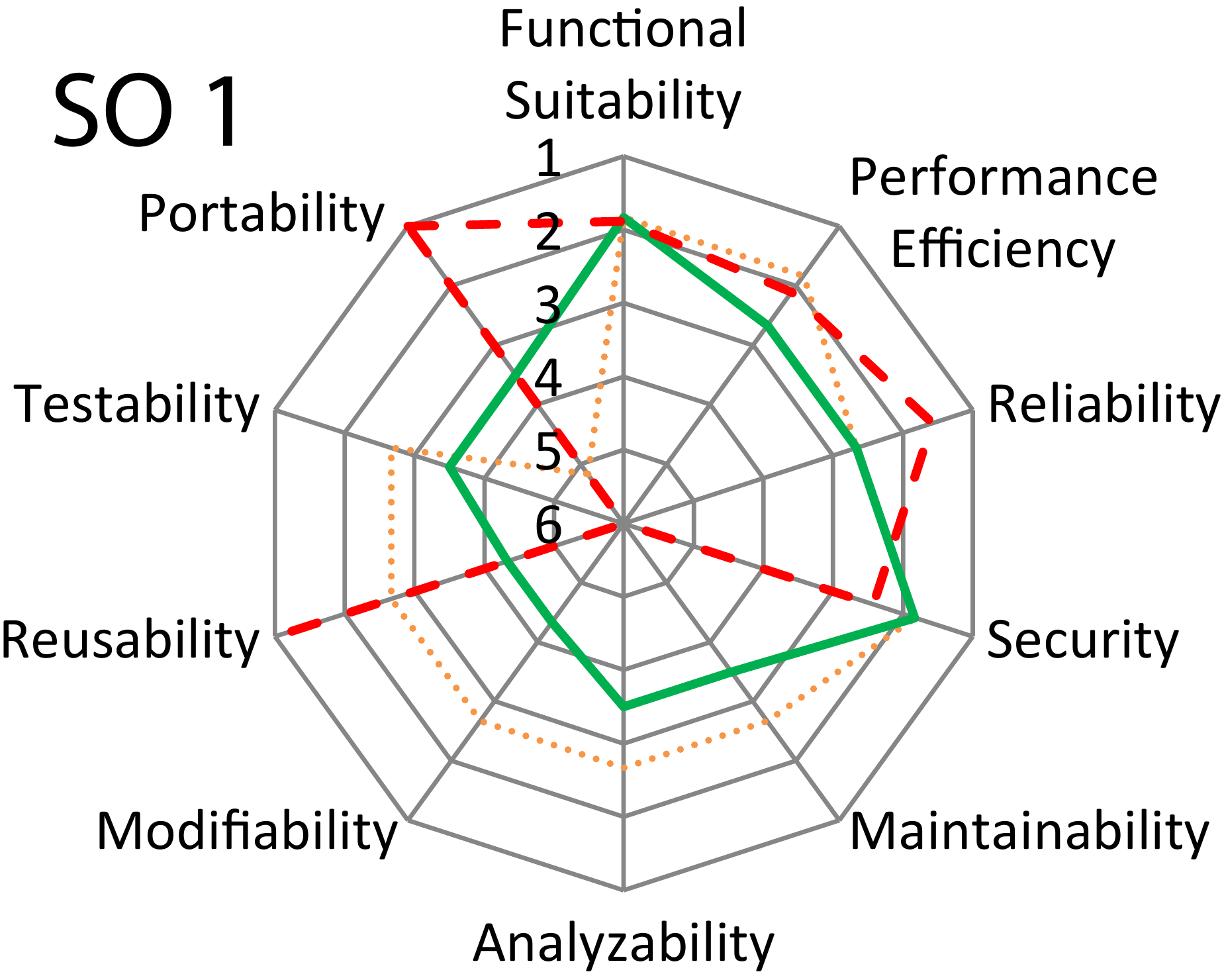} 
  }
  \subfigure[Study object 3]{
    \label{fig:kiviatRQ32SO3}
    \includegraphics[height=4cm]{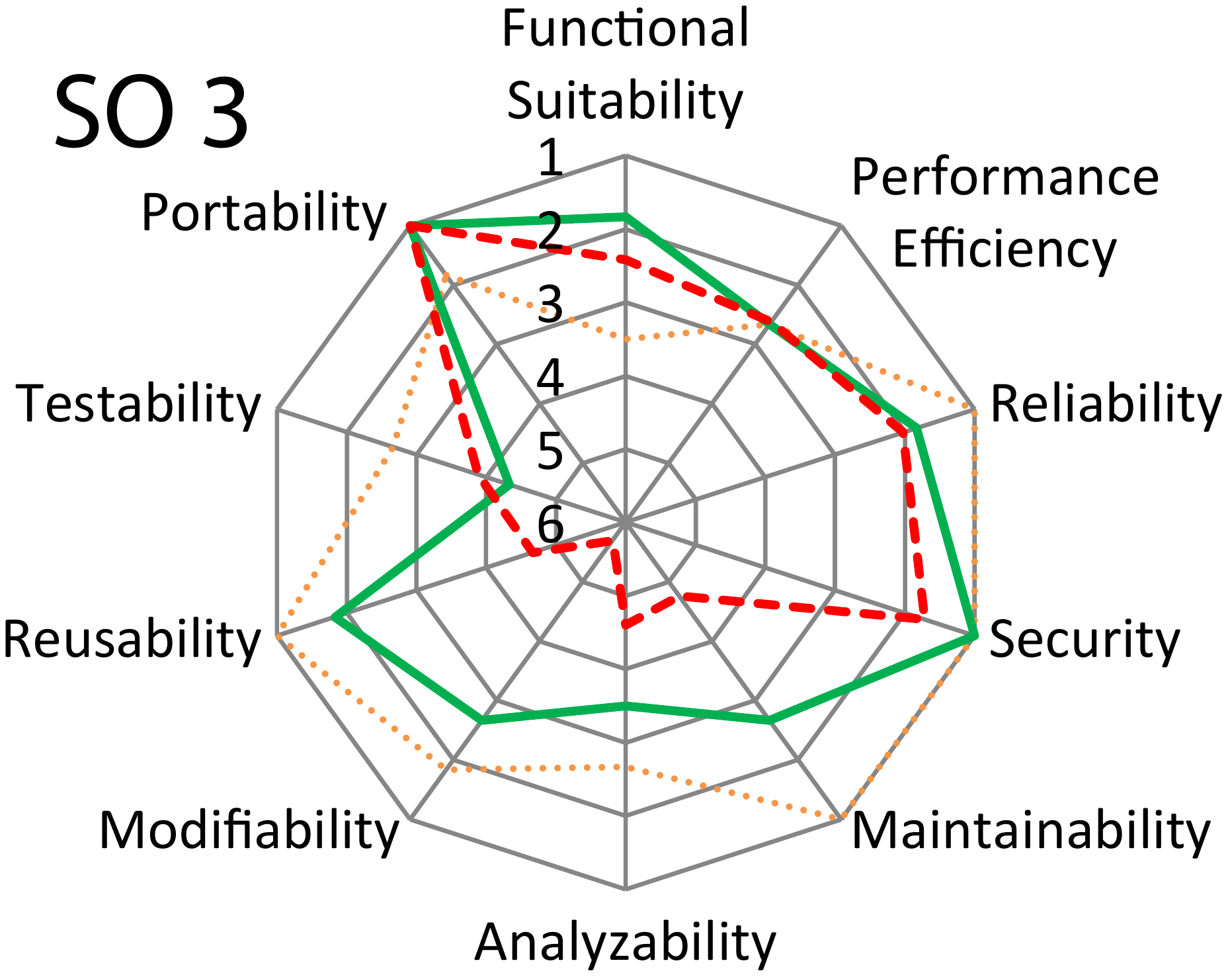} 
  }
  \subfigure[Study object 5]{
    \label{fig:kiviatRQ32SO4}
    \includegraphics[height=4cm]{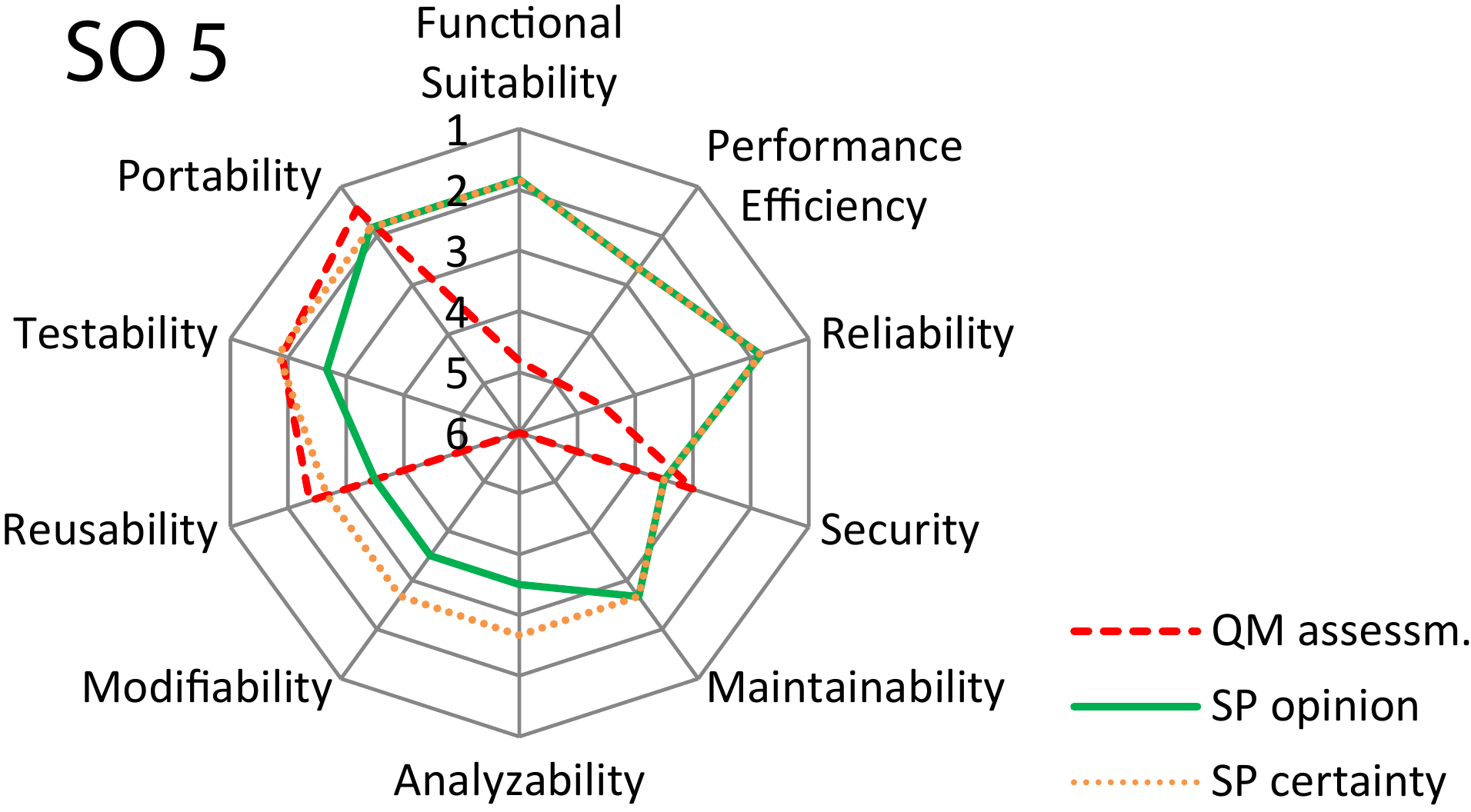} 
  }
  \caption{The result of the questionnaire in contrast to the quality model. The results were normalised according to German school grades on a scale from 6
(fail, $\stackrel{\scriptscriptstyle\wedge}{=}$ grade 6 in the quality model, level 1 in the questionnaire) to 1 (excellent, $                  \stackrel{\scriptscriptstyle\wedge}{=}$ grade 1 in the quality model, level 7 in the questionnaire).
\textit{Dashed 
line}: assessment of the quality model;
\textit{solid 
line}: opinion of the study participants according to the questionnaire; 
\textit{dotted 
line}: certainty of the study participants according to the questionnaire.}
  \label{fig:kiviatsRQ32}
\end{figure}

Following our procedure, we received three answered questionnaires from the
study participants for SO~1, 3 and 5. Their evaluation revealed differing results compared to the
grades given by the quality model. Three radar charts, 
as illustrated by Figure~\ref{fig:kiviatsRQ32}, show the estimated rating
and confidence by the study participants as well as the result of the quality model.
The three SPs rated their study object in most of the characteristics better than
it is was done by the quality model. The most interesting deviation occurs for ``maintainability'', where each
of the three study objects achieved a bad (SO~3: grade 4.76) or even the worst possible grade (SO~1 and 5: grade 6), whereas all study participants gave better ratings.

Interestingly, all of the study participants reported that they were quite confident (5 out of 7) of their
estimation. Accordingly, there is a mismatch between the ratings provided by the quality model and the opinion of the
SPs. Tables~\ref{tab:quamocoRes} and \ref{tab:rq3.2} provide further details.

\subsection{Summary and Overview of Results}
\label{sec:results:summary}

Table~\ref{tab:results:summary} depicts the most relevant findings among the results and answers to RQ 1, 2 and 3 in the previous sub-sections. This overview will be reflected and further interpreted in the following discussion.

\begin{table}
\footnotesize
\renewcommand{\arraystretch}{1.4}
\begin{tabular*}{\columnwidth}{>{\bf}L{1.9cm}|L{2.7cm}|L{2.8cm}|L{3.4cm}}
RQ & 
\textbf{Clone Detection (CD)} & 
\textbf{Bug Pattern Detection (BP)} & 
\textbf{Architecture Conformance (AC)}
\\\hline\hline
1.1\newline Technical obstacles & 
\newfinding{CD}{1.1}{1}{multiple languages and semantic clones \newline}
\newfinding{CD}{1.1}{2}{false positives by generated code}
& 
\newfinding{BP}{1.1}{1}{hard estimation of criticality and false positives\newline}
\newfinding{BP}{1.1}{2}{difficult rule selection and findings filtering}
& 
\newfinding{AC}{1.1}{1}{non-existent architecture specification (2 SOs) \newline}
\newfinding{AC}{1.1}{2}{usage of dynamic patterns (1 SO)}
\\\hline
1.2\newline Spent effort& 
\newfinding{CD}{1.2}{1}{low effort for CD \newline}
\newfinding{CD}{1.2}{2}{medium effort for gapped CD}
& 
\newfinding{BP}{1.2}{1}{high effort for aggregating data\newline}
\newfinding{BP}{1.2}{2}{needs frequent readjustment}
& 
\newfinding{AC}{1.2}{1}{high effort for understanding of specification\newline}
\newfinding{AC}{1.2}{2}{low effort of application\newline}
\newfinding{AC}{1.2}{3}{effort for application scales with number of findings\newline}
\newfinding{AC}{1.2}{4}{creation of a missing architecture specification is too costly}
\\\hline
2.1\newline Found\newline defects & 
\newfinding{CD}{2.1}{1}{two of five SOs exhibited high clone rates\newline}
\newfinding{CD}{2.1}{2}{no evidently inconsistent changes\newline}
\newfinding{CD}{2.1}{3}{large clone classes}
& 
\newfinding{BP}{2.1}{1}{few consensually critical findings in two of five SOs\newline}
\newfinding{BP}{2.1}{2}{many language-independent code complexity issues}
& 
\newfinding{AC}{2.1}{1}{four types of critical findings \newline}
\newfinding{AC}{2.1}{2}{avoidable defects in productive code}
\\\hline
2.2\newline Perceived \newline usefulness & 
\newfinding{CD}{2.2}{1}{CD was relatively new to participants\newline}
\newfinding{CD}{2.2}{2}{perceived as useful and planned to be introduced}
& 
\newfinding{BP}{2.2}{1}{already in use but should be enhanced}
& 
\newfinding{AC}{2.2}{1}{AC was new to participants \newline}
\newfinding{AC}{2.2}{2}{all partners rated AC medium to highly important and plan its application}
\\\hline
\end{tabular*}

\begin{tabular*}{\columnwidth}{>{\bf}L{1.9cm}|L{10cm}}
\\
RQ & 
{\textbf{Quality Model (QM)}}
 \\\hline\hline
3.1 Matching with ASA &

\newfinding{QM}{3.1}{2}{QM ranking is similar to critical BP and AC rankings but differs slightly from total BP ranking\newline}
\newfinding{QM}{3.1}{1}{QM ranking differs strongly from CD ranking}
\\\hline
3.2 Matching with SP & 
\newfinding{QM}{3.2}{1}{mismatch between the QM and the study participants' assessment \newline}
\\\hline
\end{tabular*}
\caption{Summary and overview of findings 1 to \arabic{findings} of RQ 1--3
\label{tab:results:summary}}
\end{table}

\section{Discussion}
\label{sec:discussion}

In the following, we discuss the results of Section~\ref{sec:results}, particularly focusing Table~\ref{tab:results:summary}
which summarizes our findings (denoted by F\emph{x}), some general observations as well as possible solutions to critical problems and lessons learned from performing our analyses.

\subsection{Reflection of Results}

Regarding RQ~1.1, we faced some obstacles during introduction and application of all ASA techniques. 
For RQ~1.2, clone detection required relatively low effort (\findref{CD}{1.2}{1}) whereas the other two techniques took significantly more time to be conducted (\findref{BP}{1.2}{1}, \findref{AC}{1.2}{1}).
Regarding RQ~2.1, we found large clone classes in SO~2 and 4 (\findref{CD}{2.1}{3}), aside from smells and pedantry a sufficient%
\footnote{Following \citet{DBLP:conf/icst/WagnerDAWS08}, three released bugs would suffice to justify ASA efforts.} 
number of pattern-based bugs in SO~1 and 5 (\findref{BP}{2.1}{1}), and 
unacceptable architecture violations in SO~1 and 4 (\findref{AC}{2.1}{1}). 
For RQ~3, the quality model assessment was different from both, the ASA results (\findref{QM}{3.1}{1}, \findref{QM}{3.1}{2}) and the opinions of the study participants.

\subsection{General Observations}

\paragraph{Observation 1: Code clone detection and architecture conformance analysis were quite new to our study participants as opposed to bug pattern detection (\findref{CD}{2.2}{1}, \findref{CD}{2.2}{2}).}

This may result from the fact that checking coding guidelines or style as well as simple bug pattern detection are standard features of modern development environments.
However, we consider it as important to know that code clone detection can indicate critical and complex relationships residing in
the code at minimum effort (\findref{CD}{1.2}{1}).
Through our work we made our SPs aware of the usefulness of architecture conformance analysis (\findref{AC}{2.2}{2}), both in the 
case of available architecture specification and the construction of a specification.

\paragraph{Observation 2: As expected, we found that all of the three techniques can be introduced and applied with resources affordable for small enterprises.}

Except for readjustment phases at project initiation or after substantial product changes, we assume that the effort of readjusting the settings for the techniques (\findref{BP}{1.2}{2}) stays very low. This effort is compensated by the time earned through narrowing results to successively more relevant findings.
Moreover, our SPs predominantly perceived the discussed and demonstrated techniques as useful (\findref{CD}{2.2}{2}, \findref{BP}{2.2}{1}) for their future projects (see Q5 in Table~\ref{tab:rq2.2_1} and Q8 in Table~\ref{tab:rq2.2_2}).

\paragraph{Observation 3: We perceived our analyses of the study objects to be successful in finding the expected maintainability issues 
as well as few critical defects (\findref{CD}{2.1}{3}, \findref{BP}{2.1}{1}, \findref{AC}{2.1}{1}).}
We share this perception with our study participants as they considered our results to be relevant (see Q3 in Table~\ref{tab:rq2.2_1}). 
In our questionnaire, we 
asked about their expectations on the results. 
The background of the SME collaboration 
and the answers to Q6 and Q7 indicate two expectations of the SPs: First, they are interested in getting a more precise idea of their SO's current quality. Second, they want to improve their capabilities, i.e.~to intensify their QA provisions, ASA tool usage and team communication. This reflects our own expectations.

\paragraph{Observation 4: Quality models provided different rankings in comparison to single ASA techniques (F27, F28).}

We found a mixed picture in comparing the rankings of single ASA techniques and the aggregated and
weighted rankings of the Quamoco quality model. Two rankings were almost the same as the quality model
ranking: Critical bug pattern findings and architecture violations. The total bug pattern ranking was similar but
ranked one SO differently and the clone detection ranking was completely different. This means that the 
aggregation in the model leads to notably different evaluation results for the SOs.
The critical bug patterns findings were ranked similar. This could mean that we considered findings
to be critical which were weighted strongly in the quality model. Assuming that the designers of the quality model created
sensible aggregations, our results suggest that it gives distinct and useful information from merely counting
findings and violations. 
We conclude that a quality model can provide a more strongly argued view on the quality of a software system.
This view can be achieved in an automated and more efficient manner which, in turn, is affordable for SMEs.

\paragraph{Observation 5: Quality models saved time in application compared to ASA techniques.}
It took us less than 90 minutes to run the analysis and to
inspect the findings. Quality models reduce the effort in comparison to the individual use of ASA
techniques, especially for bug pattern detection. This fact substantiates when taking into
account that the Quamoco tool chain performs several automated analyses and uses more rules than we had chosen for our individual analyses.

\paragraph{Observation 6: Contrary to our expectation, we partially observed a strong mismatch between the opinions of the study participants and the ratings of the quality model (\findref{QM}{3.2}{1}).}

Such a mismatch was observable although the study participants already knew of the (partially critical) defects we found by static analysis. The mismatch might stem from the fact that the quality model and the SPs apply different instruments to evaluate a
certain software system. Our quality model mainly aggregates ASA findings to assess quality characteristics: Software quality is evaluated using technical information which predominantly
resides in a system's source code. Hence, bug pattern detection is the dominating method. 

A study participant has different access to study objects and their quality assessment. The SP usually knows the course
of events throughout the SO life cycle, the measures being taken, the structure of the system, the problems that arose and the money that has been spent for different
actions. His understanding of software quality characteristics is based on intuition rather than an 
explicit quality model. Accordingly, he weighs his knowledge, tries to estimate different characteristics and reasons 
about his confidence.

In contrast, there are factors used by the quality model which the study participant may not include in his estimation or ascribe
a lesser importance or continuity of observation to them. Possible factors are e.g.\ the amount of clones in the code or the usage of default cases in switch statements. Such factors can hardly be estimated by humans without the usage of static analyses but form the basis for the quality model assessment. It is also possible that a study participant is not well prepared to process the huge amount of different aspects and static analysis warnings and, hence, neglects some information. We cannot give a clear answer to whether the quality model or the study participants achieve a better quality assessment.

\subsection{Overcoming Obstacles and Limitations in RQ~1.1 and 2.1}

Technical problems during ASA (RQ~1.1) and limitations in finding defects (RQ~2.1) are strongly governed by the effort spent and the techniques applied for
incorporating highly specific knowledge about the SOs into the analysis,
e.g.~very careful fine-tuning of the ASA tools, clone and bug-pattern filters or architecture models. In the following, we discuss some approaches overcoming these obstacles and limitations.

For code clone detection (\findref{CD}{1.1}{1}, \findref{CD}{1.1}{2}), 
\citet{Lanubile2003} performed research on finding clones in web applications usually consisting of mixtures of mark-up and procedural code. Our approach is technically limited in analysing such software. Introducing a semi-automatic approach could remove this limitation. 
\cite{DBLP:conf/csmr/DeissenboeckHHW12} discuss the challenges concerning \emph{semantic clones}.
\cite{elva2012jsctracker} sketch their dynamic detection based on what they call input-output-effect behaviour. Although their investigation of false positives lacks maturity, their detector for identifying such clones has potential to improve our results.

For bug pattern detection (\findref{BP}{1.1}{1}, \findref{BP}{1.1}{2}),
\cite{Bessey2010} confirm that misunderstood explanations of findings causes true errors to be ignored or, worse, transmuted into false positives. Thus, complicated analyses have to be well explained to developers. More than 30\% of perceived false positives give them reasons to ignore ASA tools at all.
The tools used for this report made it quite difficult and tedious to manually adjust filters to compensate for this problem.
However, scientists worked on several solutions:
For example, \citet{DBLP:conf/iwsm/FerzundAW08} report on the effectiveness of rules for smell detection. The rules they developed are based on machine learning 
and source file statistics provided by static code metrics. They used training information from two software projects including bug databases.
\cite{Kremenek2008} thoroughly discusses false positive filtering using Bayesian networks and statistical reasoning. Beyond that, he utilises specification inference to enhance detection to further types of defects.
\cite{Ruthruff2008} statistically analysed warnings of FindBugs deployed in a large software organisation. They could successfully
reduce both, (a) spurious false positive warnings (in several studies up to a third of the warnings even from sophisticated tools)
and (b) legitimate warnings without action (about half of the warnings in their study).
For a television control system, \cite{Boogerd2009} investigated the relationship between coding guideline violations detected by ASA tools
and bugs identified and treated by developers. The positive correlation 
is restricted to a small part of the applied rules (10 out of 88) and very sensitive to the project.
In the report at hand, we did not address the estimation of rule effectiveness but focused on their manual selection and application.

For architecture conformance analysis (\findref{AC}{1.1}{1}, \findref{AC}{1.1}{2}), we confirm the general limitations reported by \cite{passos2010}. Constraints depending on dynamic information can not be checked by current architecture conformance tools. Nevertheless, we think that this limit could be compensated by the use of dynamic analyses which we did not take into consideration. To the best of our knowledge this approach is not described in the literature.

In summary, utilising some of these approaches alleviates the perceived obstacles and limitations by decreasing but not necessarily eliminating the need to incorporate individual or SO-specific knowledge into the analysis.

\subsection{Usage Guidelines}

During the repetitive conduct of Steps 2 and 3 of our procedure, we gained more experience in applying the chosen techniques. For their introduction and application to a new project, we consider the following generic procedure as helpful:
\begin{enumerate}
  \item Establish a project-specific \emph{configuration}. This particularly includes the choice of bug pattern rules and the customisation of the quality model to reflect the relevant design or coding guidelines and quality characteristics.

  \item Define 
    events for \emph{measurement, findings filtering, quality assessment and for documentation}. Filtering and assessment requires in-depth knowledge of stakeholder requirements, the system and its critical components. For bug pattern detection this knowledge influences severity and confidence levels. For architecture conformance analysis this knowledge influences the definition of allowed, tolerated, and forbidden dependencies.
  
  \item Decide whether to \emph{treat or tolerate} findings and bad quality grades. This decision involves (i) the inspection of results and defective code, (ii) the issuing of change requests for defect removal and, (iii) the documentation of efforts spent to assess efficiency.

  \item Determine whether and how defects or bad quality grades can be \emph{avoided} using the lessons learned from defect treatment for future coding practice and process management.

  \item Strengthen \emph{quality gates} through improved criteria that follow patterns, such as
    ``Clone coverage in critical code package $A$ below $X\%$ prior to any bundled feature introduction.'', 
    ``No critical security errors with confidence $> Y\%$ according to tool $Z$ for any release.'',
    ``No architecture violations originating from change sets of new features.'', or ``Grade $\leq 2.0$ for analysability of core components.''

  \item For \emph{project control} in the context of \emph{continuous integration}, derive statistics and trends from findings reports by a quality control dashboard such as ConQAT or the Quamoco tool chain.
\end{enumerate}
\cite{Chandra2006} provide comparable usage guidelines for static analysis tools.

\section{Threats to Validity}
\label{sec:threats}
 
In the following, we discuss threats to the validity of our results. 
We structure them into internal and external threats.
 
\subsection{Internal Validity} 

First, a potential threat to the internal validity is that most of our project
partners had little experience with most of the applied ASA tools.
This could imply additional technical problems (RQ~1.1), which would not have
occurred when collaborating with advanced tool users or experts. Furthermore, the efforts measured for RQ~1.2 may be larger. We
mitigated this risk through discussions with experts and we assume that the
introduction in other companies would also be performed by non-experts.

Second, we did not record exact time measurements of the efforts spent.
We rather made order of magnitude estimations. We consider this threat to be small as such estimations should be sufficient in a project management context.

Third, we did not check empirically whether or how the defects, which we consensually perceived to be critical, contributed to costly system failures during operation or significant budget overruns during the SO life cycle (Section~\ref{sec:results_bugpatterns:rq22}). Due to the lack of knowledge about these life cycles, the defect criticality may have been estimated improperly. 
As discussed for the results of gapped clone detection or bug pattern detection, it is hard to properly estimate the impact of such defects.
As already mentioned, there were many easily detectable false positives. Aside from these, we tried to reduce more subtle false positives among the most relevant defects listed, through detailed inspection and group discussion.

Fourth, the questionnaire results could be wrong, due to study participants either knowingly or unknowingly giving incorrect answers. We mitigated this threat by asking them to be careful in filling it out while, at the same time, assuring anonymity to them.

Fifth, we did not check all the relationships and evaluation specifications in the
Quamoco base quality model. Hence, it is possible that there are incorrect relationships
distorting our results. We consider this threat to be small as the quality model
was inspected by several experts and showed good empirical validation results \citep{wagner:icse12}.
Furthermore, we discuss differences between the results of the quality model assessments and our
manually interpreted ASA results in RQ~3.1.

Sixth, the quality model aggregates findings in a specific way that may not be suitable for every quality perception. For example, the weighted aggregation of product factors may not take into account relative frequencies of findings. Such a normalisation leads to the effect that a few findings of one measure, normalised by a few related entities, are weighted the same as a large number of findings of another measure, even if the measures themselves have equal weights.

Seventh, we only considered popular free or open source ASA tools in our study. The problems we discussed for RQ~1.1 and in Section~\ref{sec:discussion}, in particular the difficulties in rule selection and findings filtering, may have been less severe if we used high-end commercial ASA tools 
providing additional capabilities and comfort in use. 
However, we had no budget and time available for obtaining proper licenses.
We reduced this threat by additional effort for thorough manual interpretation and reflection of all finding reports.

Eighth, we did not receive all questionnaires for RQ~3.2 from our 
project partners (3 out of 5). This lack of data could weaken our 
observation of a mismatch between the assessment of study 
participants and the quality model. We decided to include the 
results but made transparent that our interpretation relies on incomplete 
data.

Finally, in our procedure for bug pattern detection, we left out unimportant rules. This influences both, the quality model and the manual ASA interpretation. The Quamoco tool chain was only provided with the ASA tool results as an input rather than with the original code base to fully rerun these analyses. However, we do not consider this threat as a source of discrepancies in the answers to RQ~3.1.

\subsection{External Validity} 

For an experience report on a technology transfer project, the results are inherently
difficult to generalise. We had five SMEs located in Germany and four SOs specified or developed in this region.
We restricted our analysis to systems realised in Java and C\# and only applied specific analysis tools.
Hence, the problems, defects, and perceptions may be particular to this setting.

Nevertheless, we think that most of our experiences are valid for other contexts as
well. The SMEs we collaborated with range in size from only several to
one hundred employees. The domains they build software for differ strongly. Finally,
we used more popular ASA tools that had been used in industrial projects before.
However, the restriction to two programming languages has a strong effect. For other languages,
there may exist different tools and defects. For instance, with bug pattern detection,
\cite{DBLP:conf/icsea/AhsanFW09} report that characteristics 
of bug patterns can be language specific.

\section{Related Work}
\label{sec:relatedwork}

We first discuss the relevance of SMEs in our research domain followed by a survey of applications of single ASA techniques. The chapter closes with related research on quality models.

\subsection{Consideration of Software Quality in SMEs}

The work at hand enhances previous results published in \citep{Gleirscher2012-SWQD12} by investigating how quality models help in understanding data gained from static analysis.
We focus on applying static quality analysis to leverage SMEs mitigating their risks of defect-related costs. Deviating from our aim, the research community devotes its attention primarily to software process improvement (SPI) in SMEs.
\cite{Kautz2000} and \cite{Pino2008, Pino2009} report on SPI introduced in many SMEs from the late 90's till 2006 to assess software processes, change the organisations and increase their software productivity. However, they do not investigate the impact of SPI on software product quality attributes at all. 
Even the process improvements are often only measured by means of informal and non-objective processes based on SME employees' perceptions. \cite{Pino2008} conclude that SEI CMM(i) and ISO SPICE are difficult to be applied to SMEs.

\citet{Kautz1999} developed and used metrics at three SMEs to evaluate how new practices and tools for configuration and change management were affecting their software processes. Thereafter, the key to successful software measurement is making metrics meaningful and tailoring them to a particular organisation. We confirm this observation for software measurement.

\citet{Wangenheim2006} assessed software processes in SMEs and developed MARES, a set of guidelines for ISO/IEC 15504-conforming software process assessment in small companies. 
Our usage guidelines 
may form a bridge between ASA and more general guidelines for software process improvement.

\cite{hofer2002} states that only 10\% of the surveyed SMEs in the Austrian software industry believe to suffer from a lack of methods. He concludes that proper tool support as well as knowledge of methods are available. On the contrary, we argue that SMEs may not be aware of the most effective methods and can therefore not estimate their lack concerning these techniques.

\subsection{Application of ASA Techniques in SME Projects}

To the best of our knowledge, there is no study on combined automated static analysis in an SME context. However, the following publications propose and investigate single techniques separately and in different application contexts:

\paragraph{Code Clone Detection.}
\citet{Lague1997} report on application of clone detection to a large telecommunication software system. They restricted the technique to the comparison of whole methods. Opposed to that, we allowed arbitrary code fragments to be compared with each other but analysed smaller systems. Nevertheless, we confirm that clone detection can improve software maintainability.
\cite{Juergens2011} performed clone detection in different software artefacts like source code, requirements specifications and models at five enterprises. He presents a large case study investigating the impact of cloning on program correctness, an analytical cost model that quantifies the impact of code cloning on maintenance activities and a comprehensive method for clone control and tool support for practical use. 

\paragraph{Bug Pattern Detection.}
\citet{ayewah07} evaluate the accuracy and value of FindBugs and discuss but not solve the problem of properly filtering false positives. They use the term \emph{trivial bugs} 
for what we call smells and pedantry. We confirm their conclusions on the usefulness of findings and believe that an application of bug pattern detection has to undergo calibration guided by the staff of a software project.
Moreover, by answering RQ~2 with relevant findings of several tools, we contribute to Foster's, Hicks' and Pugh's (\citeyear{foster2007improving}) question ``Are the defects reported by [static analysis] tools important?''

Based on an industrial robot control system ($\approx$~2500~kLoC), \cite{springerlink:10.1007/s11219-011-9138-7} encountered drawbacks of generic ASA, i.e.\ low effectiveness, too many violations or false positives, and lack of result verifiability. They propose ASA to be tailored by system-specific rule selection to reduce false positives and ease violation handling. This includes the analysis of
annotations in comments, C macros, 
coding constraints specific to files or methods (i.e.\ call structure, control or data flow patterns), and constraints on thread scheduling or shared data usage.
Accordingly, \cite{DBLP:conf/esem/SjobergAM12} report an overrating of many maintainability metrics or predictors---often represented by bug pattern rules---and suggest more sophisticated evaluation of their system-level impacts.
For the positive impact on software quality characteristics, they conjecture: Individual rule selection is better than taking generic tool presets. They did not empirically validate this conjecture. But we qualitatively confirm it through our results on RQ~2.1 which also motivated rule selection as a cornerstone of our usage guidelines.

\citet{DBLP:conf/icst/WagnerDAWS08} similarly applied FindBugs and PMD to two industrial projects. They could not find defects reported from the field that are covered by bug pattern detection. However, our results show that this technique could indeed prevent critical defects from staying hidden until after product release.

\paragraph{Architecture Conformance Analysis.}
\citet{Rosik2008industrial} conducted an industrial case study on architecture conformance with three participating software engineers. They conclude that this technique should be integrated into the software engineering process and applied continuously. We think that the procedure we presented is able to satisfy their needs, because it explicitly focuses on continuous integration.
\citet{Mattsson2007} illustrate their experience in an industrial project and the huge effort needed to keep the architecture specification in conformance with the implementation. However, they tried to reach this goal in a manual way. Our results show that automation can reduce efforts dramatically.
\citet{Feilkas2009loss} analysed three .NET platform projects of Munich Re, a large insurance carrier, in a way similar to our procedure. In addition, they regarded loss of architectural knowledge and its effects. Compared to our results, they report a much higher effort of about five days to apply the technique primarily due to time consuming discussions. We think that the lower effort we are reporting is mainly caused by the fact that we were collaborating with small enterprises with a lower communication overhead.

\subsection{Quality Models}

To the best of our knowledge, there are no studies on applying quality models in SMEs. Hence, we describe the
relation of our study to investigations on quality standards in SMEs and practical applications of quality models.

\cite{Pusatli2011} reported on a survey with SMEs and their usage of quality standards. They found that quality is often
not a prime objective in smaller companies and especially the micro-companies do not even know the standards.

\cite{Bansiya2002} proposed QMOOD, a quality model based on object-oriented metrics. These metrics are static measures
capturing the quality of an object-oriented design. In a validation study, they found a statistically significant correlation 
between their quality index and expert opinions. They applied a much larger quality model outside an SME context.

\cite{heitlager07} developed a quality model similar to what we used in our study. Their model includes static 
maintainability measures. They discussed measurements for real systems but no further validation,
especially not in SMEs. \cite{DBLP:journals/sqj/BijlsmaFLV12} applied a maintainability model to analyse open source 
systems and compared the quality rating with defect resolution times.
They found a statistically significant correlation.
Again, we used a more detailed quality model in the SME context.

\cite{ploesch:quatic10} proposed a method for continuous code quality management using static analysis based on
a quality model. They did not prescribe a specific model but they suggested to tailor existing ones.
These authors used such a model to get different views on and understand static analysis results. Furthermore, they
reported on positive experience inside Siemens, a very large organisation, but neglected the specific situation of SMEs.

\cite{wagner:icse12} described the Quamoco model used in our study. They applied it to open source as well as commercial 
systems where they found a statistically significant correlation between quality ranking and expert opinions.
In an extended version, \cite{wagner:tse-quamoco} investigated its acceptance with software developers. The developers 
perceived the results to be reasonable and helpful for explaining the (static) measures.

In summary, there are only a few studies looking at quality models that link quality characteristics with tangible measures
and apply them in a practical setting. The existing studies derived promising correlations between expert opinions
and calculated quality evaluations. Studies applying such models to SME projects have been missing so far.

\section{Conclusions and Future Work}
\label{sec:conclusions}

It is most effective to combine several QA
methods to find most of the defects~\citep{Littlewood:2000:MEC:358134.357482}. However, this entails the cost of 
performing many different techniques. Particularly, SMEs have difficulties in assigning large
efforts to diverse techniques and training specialists for them. Automated
static analysis techniques promise to be an efficient contribution to software QA, because their repeated application requires only
little manual effort.

We reported our experience of applying three static analysis techniques to small enterprises: 
Code clone detection, bug pattern detection and architecture conformance analysis. We 
assessed potential barriers for introducing these techniques as well as the observations we could 
make in a one-year project with five German SMEs.

We found several technical problems, such as multi-language projects with single-language clone analysis 
and false positives with bug pattern analysis. However, we believe that these are no major 
road blocks for the adoption of static quality analysis. Overall, the effort for introducing the
analyses was small. Two of the techniques and the quality model could be set up with an effort of less than one person-hour.
We found various defects, such as high rates of cloning, null pointer access, erroneous calculations or circumvention of architecture layers. In the end, our study participants found all of the presented techniques relevant for inclusion into their quality assurance processes.
The use of a quality model enabled us to efficiently gain additional insights but revealed results that differed from the study participants' opinions.

In our view, static analysis tools combined with quality models can efficiently improve quality assurance in SMEs
if the techniques are continuously used throughout the development process and technically well 
integrated into the tool infrastructure. As our research was not focused on long term observations 
we can not address continuous use. Hence, it is a promising area of future work to investigate 
the long-term effects of static quality analysis on SME software projects and the seamless integration of corresponding, automated techniques into agile software development 
processes. Questions arising from the application of these techniques are ready to be examined, such as their long-term efficiency, their inclusion into an overall QA strategy, their acceptance by developers, their application to non-code development artefacts, or their effects
on daily programming. 

We are going to continue working in this area to better understand the needs of SMEs and investigate our current findings.
In particular, we will use our experiences from this and other studies to continue improving the structure of
quality models. For example, \cite{Lochmann2011} propose an extension of the Quamoco approach that
improves the support of procedural aspects of software quality. In addition, further studies are needed that investigate
the application of quality models apart from code quality assessments, for example, to the systematic derivation of software quality requirements~\citep{Lochmann2010}.

\begin{acknowledgements}
We would like to thank Christian Pfaller, Bernhard Sch\"atz and Elmar J\"urgens for their technical and organisational support throughout the project. The authors owe sincere gratitude to Klaus Lochmann for his advice and support in issues related to quality models.
We thank all involved companies as well as the OpenMRS lead developers for their reproachless collaboration and assistance.
Last but not least, we thank Veronika Bauer, Georg Hackenberg, Maximilian Junker and Kornelia Kuhle as well as our
anonymous peer reviewers for many helpful remarks.
\end{acknowledgements}

\bibliographystyle{spbasic}      

\pagebreak
\begin{appendix}
\appendix
\label{sec:appendix}

\section{Results of the two Questionnaires and the Quality Model}
\label{sec:questionnairedetails}

\setlength{\rotFPtop}{0pt plus 1fil}
\setlength{\rotFPbot}{0pt plus 1fil}
\begin{sidewaystable}
  \footnotesize
  \begin{tabular}[\textheight]{p{5.2cm}|l}
  \textbf{Question} & \textbf{Closed Answers} (without comments) \\\hline\hline
  
  \textbf{Q1)} Which of these static analysis tech\-niques have you already been using in your projects? &
  \begin{tabular}[t]{l|L{.5cm}C{.5cm}C{.5cm}C{.5cm}R{.3cm}}
    & \rb{daily} & \rb{weekly} & \rb{monthly} & \rb{less freq.} & \rb{never} \\
    Architecture conformance & 0 & 0 & 0 & 1 & 4 \\
    Bug pattern detection    & 2 & 2 & 1 & 0 & 0 \\ 
    Code clone detection     & 0 & 0 & 0 & 2 & 3 
  \end{tabular}
  \\\hline
  
  \textbf{Q2)} What is your estimate of the experience of your company in these techniques? &
    \begin{tabular}[t]{l|p{.5cm}p{.5cm}p{.5cm}p{.5cm}R{.3cm}|c}
	    & ++ & + & o & -- & - - & none \\
	    Architecture conformance & 1 & 2 & 1 & 1 & 0 & 0 \\
	    Bug pattern detection    & 1 & 3 & 1 & 0 & 0 & 0 \\
	    Code clone detection     & 0 & 0 & 1 & 0 & 1 & 3 
    \end{tabular}
  \\\hline
  
  \textbf{Q3)} How do you perceive the relevance of our analysis results for your study object? & 
	\begin{tabular}[t]{l|p{.5cm}p{.6cm}p{.55cm}p{.5cm}r|c}
	    & \multicolumn{2}{l}{high} & o & \multicolumn{2}{r|}{low} & none \\
	    Architecture conformance & 3 & & 2 & & 0 & 0 \\
	    Bug pattern detection    & 2 & & 3 & & 0 & 0 \\
	    Code clone detection     & 3 & & 2 & & 0 & 0 
	\end{tabular}
  \\\hline
  
  \textbf{Q4)} How much education could you gain from the topics of our research project? & 
	\begin{tabular}[t]{l|p{.5cm}p{.5cm}p{.5cm}p{.5cm}R{.3cm}|c}
	    & \multicolumn{2}{l}{much} & o & \multicolumn{2}{r|}{little} & none \\
	    Architecture conformance & 2 & 2 & 1 & 0 & 0 & 0 \\
	    Bug pattern detection    & 2 & 0 & 1 & 1 & 1 & 0 \\
	    Code clone detection     & 2 & 2 & 1 & 0 & 0 & 0
	\end{tabular}
  \\\hline

  \textbf{Q5)} Which of the following analysis techniques do you plan to apply at which level of priority? &
	\begin{tabular}[t]{l|p{.5cm}p{.5cm}p{.5cm}p{.5cm}R{.3cm}|c|c}
	    & ++ & + & o & -- & - - & none & *) \\
	    Architecture conformance & 1 & 3 & 0 & 1 & 0 & 0 & 5 \\
	    Bug pattern detection    & 4 & 1 & 0 & 0 & 0 & 0 & 5 \\
	    Code clone detection     & 0 & 2 & 3 & 0 & 0 & 0 & 5 
	\end{tabular} \\
	& *) application of the technique is planned
  \\\hline
 
  \end{tabular} 
  \caption{Summary of closed answers of the questionnaire for RQ 2.2 (five results, contents and answers have been \newline translated from German to English). \textbf{Legend:} ++ .. very high, + .. high, o .. medium, -- .. low, - - .. very low
  \label{tab:rq2.2_1}}
\end{sidewaystable}

\begin{table}[h!]
  \footnotesize
  \textbf{Comments and Open Answers}
  \hrule
  \vspace{.2cm}
  \textbf{Q1)}
  Architecture conformance analysis has not been used because \dots
  \begin{itemize}
    \item SS2: ``projects have been developed cleanly or without [need of] architecture.''
    \item SS1: ``manual inspection was carried through.''
    \item SS5: ``the prerequisites \dots would have needed to be established for our projects. Manual inspection (code reviews) already takes place irregularly.''
    \item SS3: ``it was not known to us.''
  \end{itemize}

  Code clone detection has not been used because \dots
  \begin{itemize}
    \item SS2: ``[clones were] not known to us as a problem.''
    \item SS3: ``we did not recognise its necessity.''
  \end{itemize}

  \vspace{.2cm}
  
  \textbf{Q3)}
  The results have been relevant because \dots
  \begin{itemize}
  \item SS1: ``manual [code] analysis is significantly more cost-intensive, \dots clone detection is only feasible with tool support.''
  \item SS5: ``we learned about concepts, experiences and tools \dots it is easier to become acquainted with [a project if its architecture conforms to its documented specification].''
  \item SS4: ``Clones are necessary within short development cycles.''
  \end{itemize}

  \vspace{.2cm}
  
  \textbf{Q5)}
	SS4: ``The results of this research project shall be included into our internal development process.''
	SS5: ``...bug pattern tools are important for early defect detection, architecture conformance and clone detection for structuring the projects.''

  \vspace{.2cm}

  \textbf{Q6)} \textit{Your estimate of the current status of your organisation w.r.t.~software quality:} 
  
  \textit{Strengths:} ``Seamless process for requirements QA \dots regarded design guidelines for all languages used (SS1) \dots flexible adaptation of guidelines to customer needs (SS1) \dots performed QA provisions (from unit testing to selective pair programming) seem to work (SS5) \dots
  so far we only experienced high customer satisfaction (SS5) \dots mature in testing techniques and management (SS3).''
  
  \textit{Weaknesses:} ``No consequent QA provisions (SS2) \dots no systematic QA (SS2) \dots automation and tool usage either project specific or even left out (SS1) \dots still learning to apply the tools (SS3).''

  \vspace{.2cm}

  \textbf{Q7)} \textit{Where do you expect the highest potential of your organisation to improve its software quality?} 
  \begin{itemize}
    \item SS2: ``Consequent QA provisions,''
    \item SS1: ``integrated tools and more automation \dots QA dashboard for project managers,''
    \item SS5: ``better knowledge transfer between teams and projects,''
    \item SS4: ``improved quality control \dots backflow of QA results into development process.''
  \end{itemize}

  \vspace{.2cm}

  \textbf{Q8)} \textit{Your estimate of the usefulness of static analysis for your software projects:}
  
  \textit{Positive:} SS2: ``Important'', SS3: ``high'', SS5: ``trend analyses are important'', SS5: ``very important, because of early and efficient defect detection \dots help identify structural deficits \dots ease [code] maintenance'', SS4: ``quality improvement starting with first build'', SS1: ``for internal projects better control and indication of deficits.''

  \textit{Negative:} SS1: ``Often not feasible in projects externally conducted at the customer's site.''
  \vspace{.2cm}
  \hrule
  \caption{Summary of comments and open answers of the questionnaire for RQ~2.2 (five results, contents and answers have been translated from German to English, SS \dots study subject, except from SS3, SSx corresponds to SOx)
  \label{tab:rq2.2_2}}
\end{table}

\begin{landscape}
\footnotesize 

\begin{longtable}{%
>{\raggedright\itshape}L{9.5cm}%
|L{6cm}%
|!{\extracolsep{\fill}}L{3.7cm}}

\caption{Quality model results matched with individual ASA results for RQ~3.1 (worst three characteristics focused, measures are ordered by their weighted impact).
  \textbf{Legend:} CD \dots Code Clone Detection, BP \dots Bug Pattern Detection, AC \dots Architecture Conformance Analysis, QM \dots quality model. Grades are in brackets (1 \dots best, 6 \dots worst).
  \label{tab:rq3.1}}\\
  
\textbf{Most Affected Quality Characteristics} \newline as graded by Quamoco quality model &
\textbf{Impression and Justification} \newline 
based on noticeable results from RQ 2.1
& \textbf{Observation}
\\\hline\hline
\endfirsthead
\multicolumn{3}{r}{(continued on next page)}
\endfoot
\endlastfoot

\textbf{Most Affected Quality Characteristics} \newline as graded by Quamoco quality model &
\textbf{Impression and Justification} \newline 
based on noticeable results from RQ 2.1
& \textbf{Observation}
\tabularnewline \hline\hline
\endhead

  \multicolumn{2}{l}{\textbf{SO 1 
  }}\\\hline
  \textbf{Security} (2.42)
  influenced by
  (BP) 
  avoid long parameter list
  &
  (BP)
  visible constants,
  constructor calls overwritable method;
  (AC) critical violations
  &
  Different rules and AC, similar impression  
  \\\hline

  \textbf{Maintainability} (6.0)
  influenced by
  (BP)
  avoid ref and out parameters, 
  avoid message chain, 
  avoid properties without get accessor,
  avoid unneeded calls on string,
  avoid unnecessary specialisation,
  avoid long parameter list, 
  avoid redundancy in method name;
  (ConQAT)
  overly long file, 
  clone coverage
  &
  (BP)
  empty/general exception handlers,
  nested use of generic types,
  deep namespaces;
  (ConQAT)
  	 block depth;
  (CD) normal clone coverage;
  (AC) critical violations
  &
  Different rules,
  QM provides more severe impression than tools directly
  \\\hline

  \textbf{Performance Efficiency} (2.11)
  influenced by 
  (BP) 
  avoid unneeded calls on string
  &   
  RQ 2.1 offered no relevant findings for this 
  &
  QM provides new impression
  \\\hline

  \multicolumn{2}{l}{\textbf{SO 2} 
  }\\\hline

  \textbf{Performance Efficiency} (1.84)
  influenced by 
  (BP)
  improper use of ``synchronized''
  & 
  RQ 2.1 offered no indication to evaluate this
  & 
  QM provides new impression
  \\\hline

  \textbf{Maintainability} (2.7)
  influenced by 
  (BP) 
  improper use of ``synchronized'', 
  unnecessary specialisation;
  (ConQAT) 
  long file, 
  duplication
  & 
  (BP) 
  empty/general exception handlers, 
  nested use of generic types, 
  deep namespaces;
  (ConQAT) 
  nested block depth;
  (CD) 
  bad clone coverage
  & 
  Different rules,
  QM provides similar impression than tools directly
  \\\hline

  \multicolumn{2}{l}{\textbf{SO 3 
  }}\\\hline

  \textbf{Functional Suitability} (2.4)
  influenced by 
  (BP)
  covariant compare method, 
  static date format,
  field is mutable hash table,
  superclass uses subclass during init,
  unused field
  & 
  (BP) 
  constructor calls overwritable method, 
  corrupted serialisable, 
  return values not validated
  & 
  Different rules,
  QM provides similar impression than tools directly
  \\\hline

  \textbf{Maintainability} (4.76)
  influenced by
  (ConQAT) 
  missing comments;
  (FindBugs)
  superclass uses subclass during init,
  public hashtable declared as constant,
  field is mutable hash table,
  use of unhashable class,
  bad casts,
  mutable hash table,
  reference to mutable object,
  covariant compare method,
  superclass names;
  (CD) 
  duplication
  &
  (BP) 
  unused local variables, 
  empty method in abstract class, 
  cyclomatic complexity;
  (ConQAT)
  nested block depth;
  (CD) normal clone coverage
  & 
  Different rules,
  QM provides a similar impression less dependent on frequency of findings
  \\\hline
  
  \textbf{Performance Efficiency} (2.6)
  influenced by 
  (BP) 
  unused field, 
  string concatenation, 
  redundant null check,
  unused local variable 
  & 
  (BP) 
  inefficient string manipulation
  & 
  More rules in QM, but similar impression
  \\\hline

  \multicolumn{2}{l}{\textbf{SO 4}
  }\\\hline
  \textbf{Performance Efficiency} (2.12)
  influenced by 
  (BP)
  unused Field, 
  redundant null check, 
  uncalled method,
  unused local variable 
  & 
  (AC) 
  critical violations
  & 
  RQ 2.1 suggests less severe impression due to ignored rules but regards AC
  \\\hline

  \textbf{Maintainability} (4.4)
  influenced by 
  (ConQAT) 
  improper comments, 
  clone coverage;
  (BP)
  field should be package protected, 
  reference to mutable, 
  questionable cast, 
  duplicate branches, 
  default case missing
  & 
  (BP) 
  extensive class/method size or parameter count or too many fields,  
  cyclomatic complexity;
  (ConQAT)
  nested block depth;
  (CD) extreme clone rate;
  (AC) critical violations
  & 
  Different rules, RQ 2.1 suggests less severe impression but regards AC
  \\\hline

  \multicolumn{2}{l}{\textbf{SO 5} 
  }\\\hline

  \textbf{Functional Suitability} (4.8)
  influenced by 
  (BP) 
  improper null parameter, 
  return value ignored,
  bitwise OR of signed byte,
  no suitable constructor,
  bad field store, 
  call to static date format,
  useless substring, 
  redundant null check, 
  useless control flow
  & 
  (BP) 
  access of a null pointer, 
  integer shift beyond 32 bits;
  (CD) normal clone rate
  & 
  QM provides much more arguments than RQ 2.1
  \\\hline
  
  \textbf{Performance Efficiency} (4.95)
  influenced by 
  (BP) 
  improper use of removeAll, 
  useless substring, 
  known null value, 
  use of inefficient iterator,
  unneeded boxing of value,
  local double assignment,
  redundant null check, 
  useless control flow,
  useless garbage collection,
  unused local variable   
  & 
  RQ 2.1 summarised no indication to evaluate this
  & 
  QM provides new impression
  \\\hline
  
  \textbf{Maintainability} (6)
  influenced by
  (ConQAT) 
  improper comments, 
  duplication;
  (BP) 
  synchronisation on boolean,
  improper use of exceptions,
  thread started by run, 
  mutable object as constant,
  field should be package protected,
  mutable hash table,
  exposure of internals,
  bitwise OR of signed byte,
  initialisation circularity, 
  use of inefficient iterator,
  call to static date format, 
  inconsistent synchronisation,
  questionable cast,
  improper use of removeAll, 
  duplicate branches,  
  useless substring, 
  known null value, 
  redundant null check, 
  useless control flow,
  improper reference to mutable,
  unrelated types,
  default case missing
  &
  (BP) 
  extensive class/method size or parameter count or too many fields,
  cyclomatic complexity,
  empty method in abstract class;
  (ConQAT)
  nested block depth;
  (CD) 
  normal clone rates
  &
  QM provides more arguments than RQ 2.1 to conclude a similarly severe grade
  \\\hline

\end{longtable}

\normalsize
\end{landscape}

\begin{table}[t!]
\begin{center}
  \renewcommand{\arraystretch}{1.1}
  \begin{tabular*}{\textwidth}{L{4cm}|c|c|c|c|c}
 \textbf{Quality Characteristic} & 
  \textbf{SO 1} &
  \textbf{SO 2} &
  \textbf{SO 3} &
  \textbf{SO 4} & 
  \textbf{SO 5}
  \\\hline\hline
  \textbf{Overall Quality} &
  [2.6-3.0] & 
  [1.3-1.7] & 
  [2.4-2.6] & 
  [1.9-2.1] & 
  [4.4-4.6] 
  \\\hline
  \textbf{Functional Suitability} &
  [1.6-2.2] & 
  [1.0-1.5] & 
  [2.3-2.5] & 
  [1.7-1.9] & 
  [4.7-4.9]   
  \\\hline
  \textbf{Performance Efficiency} &
  [1.8-2.5] & 
  [1.5-2.2] & 
  [2.4-2.9] & 
  [1.9-2.4] & 
  [4.7-5.2]   
  \\\hline
  \textbf{Reliability} &
  1.6         & 
  1.1         & 
  2.0         & 
  1.2         & 
  4.6          
  \\\hline
  \textbf{Maintainability} &
  6.0         & 
  [2.2-3.2] & 
  [4.5-5.0] & 
  [4.2-4.6] & 
  6.0           
  \\\hline
  \textbf{Analysability} &
  6.0         & 
  [3.2-5.0] & 
  [4.0-5.1] & 
  [5.0-6.0] & 
  6.0           
  \\\hline
  \textbf{Reusability} &
  1.0         & 
  1.0         & 
  4.7         & 
  1.0         & 
  2.4           
 \\\hline
  \textbf{Modifiability} &
  6.0         & 
  2.7         & 
  5.7         & 
  6.0         & 
  6.0           
  \\\hline
  \textbf{Testability} &
  6.0        & 
  [1.0-2.8] & 
  [3.7-4.1] & 
  [1.0-1.5] & 
  [1.7-2.1]   
  \\\hline  
  \textbf{Security} &
  2.4         & 
  1.0         & 
  1.7         & 
  1.4        & 
  3.0           
  \\\hline
  \textbf{Portability} &
  1.0         & 
  1.0         & 
  1.0         & 
  1.0         & 
  1.5           
  \\\hline
  \end{tabular*} 
  \caption{Results of the quality model for RQ~3.1, rounded to one decimal; rating given in German school grades - 1: excellent, 6: insufficient. Intervals indicate that manual reviews are missing, which are required by the quality model to state more precise results.
  \label{tab:quamocoRes}}
\end{center}
\vspace{-10pt}
\end{table}

\begin{table}[t!]
\begin{center}
  \renewcommand{\arraystretch}{1.1}
  \begin{tabular*}{\textwidth}{l|cc|cc|cc}
  \multirow{2}[0]{*}[-0.5ex]{\textbf{Characteristics}} &
  \multicolumn{2}{c|}
  {\textbf{Study object 1}} 
   &
  \multicolumn{2}{|c|}
  {\textbf{Study object 3}}
   &
  \multicolumn{2}{c}
  {\textbf{Study object 5}}\\
  & \textbf{Estim.} &
    \textbf{Conf.} &
    \textbf{Estim.} &
    \textbf{Conf.} &
    \textbf{Estim.} &
    \textbf{Conf.}
  \\\hline\hline
  \textbf{Functional Suitability} & 6 & 6 & 6 & 4 & 6 & 6 \\\hline
  \textbf{Performance Efficiency} & 5 & 6 & 5 & 5 & 5 & 5 \\\hline
  \textbf{Reliability}  			& 5 & 5 & 6 & 7 & 6 & 6 \\\hline
  \textbf{Maintainability}  		& 4 & 5 & 5 & 7 & 5 & 5 \\\hline
  \textbf{Analysability}  		& 4 & 5 & 5 & 7 & 5 & 5 \\\hline
  \textbf{Modifiability}  		& 3 & 5 & 5 & 6 & 4 & 5 \\\hline
  \textbf{Reusability} 			& 3 & 5 & 6 & 7 & 4 & 5 \\\hline
  \textbf{Testability}  			& 4 & 5 & 3 & 5 & 5 & 6 \\\hline  
  \textbf{Security}  				& 6 & 6 & 7 & 7 & 5 & 5 \\\hline
  \textbf{Portability}  			& 4 & 2 & 7 & 6 & 6 & 6 \\\hline
  \end{tabular*} 
  \caption{Results of the questionnaire \textit{Comparison between the results of the quality model and the study participants' opinions} (RQ~3.2). The quality estimation and confidence can range from 1 (Insufficient/Unconfident) to 7 (Excellent/Confident).
  \label{tab:rq3.2}}
\end{center}
\vspace{-10pt}
\end{table}

\end{appendix}

\end{document}